\documentclass[fleqn,usenatbib]{mnras}

\usepackage{newtxtext,newtxmath}

\usepackage[T1]{fontenc}

\DeclareRobustCommand{\VAN}[3]{#2}
\let\VANthebibliography\thebibliography
\def\thebibliography{\DeclareRobustCommand{\VAN}[3]{##3}\VANthebibliography}

\usepackage{graphicx}	
\usepackage{amsmath}	
\usepackage{mhchem}
\usepackage{todonotes,xcolor}

\newcommand{\lpup}{L$_2$\,Pup}

\definecolor{orange}{rgb}{0.82, 0.1, 0.26}

\title[A molecule-rich disk around L$_2$\,Pup]{Modelling predicts a molecule-rich disk around the AGB star L$_2$\,Puppis}

\author[M. Van de Sande et al.]{
M. Van de Sande,$^{1,2}$\thanks{E-mail: mvdsande@strw.leidenuniv.nl}
C. Walsh,$^{1}$
T. Danilovich,$^{3,4,5}$
F. De Ceuster,${^5}$
and T. Ceulemans$^{5}$
\\
$^{1}$School of Physics and Astronomy, University of Leeds, Leeds LS2 9JT, UK\\
$^{2}$Leiden Observatory, Leiden University, P.O. Box 9513, 2300 RA Leiden, The Netherlands\\
$^{3}$School of Physics \& Astronomy, Monash University, Wellington Road, Clayton 3800, Victoria, Australia\\
$^{4}$ARC Centre of Excellence for All Sky Astrophysics in 3 Dimensions (ASTRO 3D), Clayton 3800, Victoria, Australia \\
$^{5}$Institute of Astronomy, KU Leuven, Celestijnenlaan 200D, 3001 Leuven, Belgium\\
}

\date{Accepted XXX. Received YYY; in original form ZZZ}

\pubyear{2015}

\begin{document}
\label{firstpage}
\pagerange{\pageref{firstpage}--\pageref{lastpage}}
\maketitle

\begin{abstract}
The nearby oxygen-rich AGB star \lpup\ hosts a well-studied nearly edge-on disk.
To date, disks around AGB stars have not been chemically studied in detail.
By combining a parameterisation commonly used for protoplanetary disks and archival ALMA observations, we retrieved an updated density and temperature structure of this disk.
This physical model was then used as input to the first chemical model of an AGB disk. 
The model shows that the physical structure of the disk has a large impact on its chemistry, with certain species showing large changes in column density relative to a radial outflow, indicating that chemistry could be used as a tracer of disks that cannot be directly imaged.
Despite its oxygen-rich nature, the daughter species formed within the disk are surprisingly carbon-rich.
Two chemical regimes can be distinguished: cosmic-ray induced chemistry in the midplane and photochemistry induced by the interstellar radiation field in the outer regions.
Certain complex organic molecules are formed in the midplane. 
This occurs via gas-phase chemistry only, as the disk is too warm for dust-gas chemistry.
The photochemistry in the outer regions leads to the efficient formation of (long) carbon-chains. 
The predictions of the model allow us to tentatively put the disk's age $\lesssim 10^5$ yr.
Additional observations are necessary to better constrain the physical structure of \lpup's disk and are essential to test the predictions made by the chemical model.
Our exploratory work paves the way for a more general study of the chemistry of AGB disks. 
\end{abstract}




\section{Introduction}			\label{sect:introduction}

The asymptotic giant branch (AGB) phase near the end of the lives of stars with an initial mass up to 8 M$_\odot$ is characterised by vigorous mass loss.
AGB stars lose their outer layers via a stellar wind or outflow at a rate between $10^{-8}$ and $10^{-4}$ M$_\odot$ yr$^{-1}$, creating an expanding circumstellar envelope (CSE, \citealt{Habing2003,Hofner2018}).
CSEs are rich astrochemical environments, with close to 100 different molecules and some 15 types of newly formed dust grains detected so far \citep{Decin2021}.
Spherically symmetric outflows appear to be the exception rather than the rule.
Asymmetrical structures are ubiquitously observed, ranging from small-scale asymmetries, such as density-enhanced clumps \cite[e.g.,][]{Leao2006,Khouri2016,VelillaPrieto2023}, to large-scale structures, such as spirals \cite[e.g.,][]{Mauron2006,Maercker2012} and disks \cite[e.g.,][]{Kervella2016}.
These large-scale structures are thought to be caused by binary interaction with (sub)stellar companions.
Circumbinary disks are expected around younger, lower mass-loss rate AGB stars, evolving into spirals as the mass-loss rate increases and the companion's orbit widens during the AGB phase \citep{Decin2020}.

The classic spherically symmetric one-dimensional chemical kinetics model of the CSE is hence applicable to only a minority of outflows.
The effects of the UV field of a stellar companion and a clumpy substructure on the chemistry throughout the outflow have been included in chemical models of the CSE \citep{VandeSande2018,VandeSande2022,VandeSande2023}. 
These models show that chemistry can be used to unveil a hidden stellar companion \citep{Siebert2022} and retrieve its orbital properties \citep{Danilovich2024}.
The influence of large-scale density structures on the CSE's chemistry requires specialised chemical models.
So far, models have been developed for the density-enhanced shells in the outflow of IRC+10216 \citep{Brown2003,Cordiner2009,Agundez2017}, but no general models applicable to a larger set of AGB stars exist. 
The chemistry of disks around AGB stars has not been studied in detail.

The best-studied AGB disk is that around the oxygen-rich AGB star \lpup, the second-closest AGB star at a distance of 64 pc \citep{vanLeeuwen2007}.
Optical and infrared observations have shown the clear presence of an almost edge-on compact dusty disk with an inner dust rim at 6 au, spanning to around 15 au.
Loops and plumes extending above the disk have also been identified \citep{Ireland2004,Kervella2014,Lykou2015,Ohnaka2015,Kervella2015}.
ALMA observations probed the gaseous disk, which has an inner rim at 2 au, and revealed a tentative planetary companion with a mass of $12 \pm 16$ M$_\mathrm{J}$. 
This planet is probably not formed within the disk, but is instead rather part of \lpup's (first generation) planetary system \citep{Chen2016,Kervella2016}.
The physical structure of the disk has been proposed to be similar to that of protoplanetary disks (PPDs), albeit with a higher temperature in the midplane as it is assembled from warm stellar outflow material \citep{Homan2017}.

In this paper, we develop a two-dimensional chemical model of \lpup's disk: the first such model of a disk around an AGB star. 
This exploratory work probes the unique astrochemical regime of a hot and dense disk, irradiated by the interstellar radiation field.
By determining the effects of the physical disk structure on the chemistry around \lpup, we can identify the chemical differences with a spherically symmetric O-rich outflow and establish whether the parent species are redistributed into different carriers.
Such differences can be used as chemical tracers of disks that are not as easily resolved as the close-by \lpup.
Additionally, the evolution of the disk's chemical structure over time can be used to determine the age of the disk. 

In Sect. \ref{sect:physics}, we use archival ALMA data to retrieve the disk’s density and temperature structure using a parametric model commonly used for PPDs.
This physical structure is then used as input to the chemical model in Sect. \ref{sect:chem}, where we also show the chemical modelling results.
The physical and chemical modelling results are discussed in Sect. \ref{sect:disc}, followed by our conclusions in Sect. \ref{sect:conclusions}.

\section{Physical model}			\label{sect:physics}

\lpup\ was observed by ALMA during Cycle 3 for project 2015.1.00141.S (PI Pierre Kervella) in November 2015.
The reduction of the Band 7 observations is described in \citet{Kervella2016}, and we use that dataset in the present work. 
The resulting data have an angular resolution better than 15 mas and a maximum resolvable scale of 200 mas. 
The $J = 3-2$ lines of \ce{^{12}CO} and \ce{^{13}CO} from the dataset are presented in \citet{Homan2017}.
The data cubes have a spectral resolution of 0.22 km s$^{-1}$ and an rms noise level of 2.5 and 2.6 mJy beam$^{-1}$, respectively.
The channel maps of the lines are presented here in Appendix \ref{app:alma}, now with beam information.

\citet{Homan2017} derived a physical model of \lpup's disk from the \ce{^{12}CO} and \ce{^{13}CO} $J = 3-2$ lines. 
The model from that work, although provided by the first author, was unable to be reproduced.
Given the similarity of the disk to PPDs, we opted to remodel the data using a parametric model commonly used for PPDs.

The parametric model is described in Sect. \ref{subsect:physics:ppd}.
Sect. \ref{subsect:physics:magritte} describes the  3D line radiative transfer code used, \textsc{Magritte}.
The parameter space explored is laid out in Sect. \ref{subsect:physics:params}, the retrieved physical structure of the disk is presented in Sect. \ref{subsect:physics:results}.

\subsection{Parametric model of a protoplanetary disk}			\label{subsect:physics:ppd}

We use the parametrisation of \citet{Williams2014}, which assumes a disk surface density profile based on the solution for an axisymmetric, steady-state viscous accretion disk in hydrostatic equilibrium \cite[e.g.,][]{LyndenBell1974,Hartmann1998}.
The surface density of the disk is described by
\begin{equation}
	\Sigma(r) = \Sigma_0 \left( \frac{r}{r_0} \right)^{-\gamma} \ \exp\left( - \left(\frac{r}{r_0} \right)^{2-\gamma} \right),
\label{eq:sigma}
\end{equation}
where $r_0$ is a disk reference radius and $\gamma$ describes the radial dependence.  
$\Sigma_0$ gives the global normalisation to the gas mass at $r_0$, written as
\begin{equation}
	\Sigma_0 = (2-\gamma) \ \frac{M_\mathrm{gas}}{2 \pi r_0^2} \ \exp \left( \frac{r_\mathrm{in}}{r_0} \right)^{2-\gamma},
\end{equation}
with $M_\mathrm{gas}$ the disk's total gas mass and $r_\mathrm{in}$ its inner radius.
Assuming hydrostatic vertical balance in the disk, the density structure of the disk is described by 
\begin{equation}
	\rho(r,z) = \frac{\Sigma(r)}{H_p(r)\ \sqrt{2\pi}} \exp \left(-\frac{z^2}{2H_p(r)^2}\right), 
\label{eq:rho}
\end{equation}
with $H_p$ the pressure scale height derived from the midplane temperature.
This scale height is given by the ratio between the speed of sound, $c_s$, and the Keplerian angular frequency $\Omega_K$, or,
\begin{equation}
	H_p = \frac{c_s}{\Omega_K} = \sqrt{\frac{k_B\ T_\mathrm{mid}(r)\ r^3}{G\ M_*\ \mu m_H}}.
\label{eq:H}
\end{equation}
Here, $M_*$ is the stellar mass, $k_B$ is the Boltzmann constant, $G$ the gravitational constant, $m_H$ the mass of the hydrogen atom, and $\mu$ the mean molecular mass.

The midplane temperature profile is parametrised by a power law,
\begin{equation}	\label{eq:Tmid}
	T_\mathrm{mid}(r) = T_\mathrm{mid,1} \left( \frac{r}{1 \mathrm{au}} \right)^{-q},
\end{equation}
with $T_\mathrm{mid,1}$ the temperature at 1 au and $q$ the power-law exponent.
Similarly, the atmospheric temperature surrounding the disk is parametrised by
\begin{equation}	\label{eq:Tatm}
	T_\mathrm{atm}(r) = T_* \left( \frac{r}{1 \mathrm{au}} \right)^{-\epsilon},
\end{equation}
with $T_*$ the stellar temperature and $\epsilon$ the power-law exponent. 
The vertical temperature profile connects Eqs (\ref{eq:Tmid}) and (\ref{eq:Tatm}) by a sine function:
\begin{equation}
T(r,z) =
  \begin{cases}
  T_{\rm mid} + (T_{\rm atm} - T_{\rm mid})
              \left[\sin\left(\frac{\pi z}{2z_q}\right)\right]^{2\delta}
              & \mbox{if } z < z_q\\
  T_{\rm atm} & \mbox{if } z\geq z_q
  \end{cases},
\label{eq:temperature}
\end{equation}
where $\delta$ describes the steepness of the temperature profile and $z_q$ describes the height at which the disk reaches the atmospheric value.
Following \citet{Williams2014}, we fix $\delta = 2$ and $z_q = 4 H_p(r)$.
 
For a given stellar mass, the physical structure of the disk is described by eight parameters: four for the density profile, $\{\gamma, r_\mathrm{in}, r_0, M_\mathrm{gas}\}$, and four for the temperature profile, $\{T_\mathrm{mid,1}, T_\mathrm{atm,1}, q, \epsilon\}$.

\subsection{\textsc{Magritte}: 3D line radiative transfer}			\label{subsect:physics:magritte}

\textsc{Magritte}\footnote{Available at \url{https://github.com/Magritte-code/Magritte}, we used version 0.3.2.} is an open-source 3D non-LTE line radiative transfer code, mainly used for creating synthetic spectral line observations \citep{DeCeuster2020a,DeCeuster2020b,DeCeuster2022}.
We initialise the \textsc{Magritte} model by defining the model on a uniform $60 \times 60 \times 30$ au grid. 
For each point, we define the temperature, velocity, turbulent velocity, \ce{H2} density and CO density. 
The grid is then resampled to speed up computation by limiting the allowed variation of the density, which we evaluated at the uniform grid positions (Ceulemans et al., in prep.). 
After model setup,  the radiation field is computed assuming non-LTE by taking into account the first seven rotational CO lines. 
In this way, the energetic state of the CO gas is determined self-consistently. 
Afterwards, we create spectral line images for the $^{12}$CO and $^{13}$CO $J = 3-2$ lines.

\subsection{Parameter space of the physical model}			\label{subsect:physics:params}

Table \ref{table:modelparams} shows the parameters of the physical model along with the values which best reproduce the integrated emission profiles.
Based on previous observations, we fixed the inner radius of the disk, velocity profile, temperature profile of the disk's atmosphere, the fractional abundance of \ce{^{12}CO}, and the stellar and disk mass. 
These parameters are listed in the top rows of Table \ref{table:modelparams}.
The stellar radius is fixed at 1 au.
The inner radius of the disk, $r_\mathrm{in}$, is fixed at 2 au \citep{Kervella2016}.
The velocity profile is assumed to be Keplerian from the inner radius out to the inner rim of the dusty disk at 6 au. 
Beyond 6 au, the velocity profile is sub-Keplerian with a power-law exponent of 0.853 \citep{Kervella2016}.
The turbulent velocity is fixed to 0.5 km s$^{-1}$ \citep{Homan2017}.
The temperature profile of the atmosphere is fixed at $T_\mathrm{atm,1} = 2800$ K with a power-law exponent of 0.65 \citep{Danilovich2015}.
The fractional abundance of \ce{^{12}CO} was taken to be $1 \times 10^{-4}$ with respect to \ce{H2} \citep{Homan2017}.
The total gas mass of the disk, $M_\mathrm{gas}$, is fixed at $2.5 \times 10^{-4}$ M$_\odot$ following \citet{Kervella2015} and assuming a dust-to-gas mass ratio of 0.01.
The stellar mass, $M_*$, is fixed to $0.659\ \times$ M$_\odot$, as derived by \citet{Kervella2016}.

The other parameters were allowed to vary and are listed in the bottom rows of Table \ref{table:modelparams}.
Their starting points were based on the model of \citet{Homan2017}.
The midplane temperature, $T_\mathrm{mid,1}$, was varied in increments of 100 K, the power-law of the midplane temperature, $q$, in increments of 0.05, the characteristic disk radius, $r_0$, in increments of 1 au, the parameter determining the surface density (Eq. \ref{eq:sigma}), $\gamma$, in increments of 0.5, and the \ce{^{13}CO} abundance in increments of $1 \times 10^{-5}$ w.r.t. \ce{H2}.

\subsection{Physical structure of the disk}			\label{subsect:physics:results}

The model calculation was constrained by comparing slices along the disk midplane through the moment 0 (integrated intensity) maps of the observed lines. 
The moment 0 maps are presented in \citet{Kervella2016} (their Figs. A1 and B1).
The slices were chosen so that they go through the emission peaks in the East and West sides of the disk as the disk's position angle is not zero.
We choose to use this metric as a radial slice along the major axis of the disk will capture well the radial behaviour of the emission without any complication that may arise from deprojection of the emission from this highly inclined disk.
The slice through the moment 0 map of \ce{^{12}CO} is centred on (07$^\mathrm{h}$13$^\mathrm{m}$32$\fs$477, -44$^\circ$38$\arcmin$17$\farcs$845) at an angle of $9.53^\circ$, that through the \ce{^{13}CO} moment 0 map is centred on (07$^\mathrm{h}$13$^\mathrm{m}$32$\fs$475, -44$^\circ$38$\arcmin$17$\farcs$843) at an angle of $16.0^\circ$.
The slices are different as the \ce{^{12}CO} and \ce{^{13}CO} probe different layers within the disk.

We first modelled the \ce{^{12}CO} line by reproducing the slice through its moment 0 map.
Good models were determined by minimising the chi-squared value of the fit to the slice. 
For these models, we then calculated the corresponding \ce{^{13}CO} model, varying the \ce{^{13}CO} abundance as an additional parameter while keeping the others fixed and calculated the chi-squared value of the fit to the slice through the \ce{^{13}CO} moment 0 map.
The \ce{^{12}CO}/\ce{^{13}CO} ratio is fixed throughout the disk.

Our best-fitting model provides the best reproduction of the \ce{^{12}CO} moment 0 slice, with a chi-squared value of 3.92.
A \ce{^{13}CO} abundance of $2 \times 10^{-5}$ results in the fourth-best fit in our sampled parameter space to the \ce{^{13}CO} moment 0 slice, with $\chi^2 = 1.76$.
A larger \ce{^{13}CO} abundance of $3 \times 10^{-5}$ results in the third-best model, with $\chi^2 = 1.74$.
However, this would yield a \ce{^{12}CO}/\ce{^{13}CO} ratio of 3.33.
Since this ratio is lower than expected from observations \citep{Ramstedt2014} and the improvement to reproducing the observations is marginal, we chose the model with a \ce{^{13}CO} abundance of $2 \times 10^{-5}$ as our best model.
Note that the model that best reproduces the \ce{^{13}CO} moment 0 slice, with the lowest chi-squared value of 1.65, corresponds to the fourth-best model to the \ce{^{12}CO} moment 0 slice, with $\chi^2 = 4.19$. 

\begin{figure*}
 \includegraphics[width=0.8\textwidth]{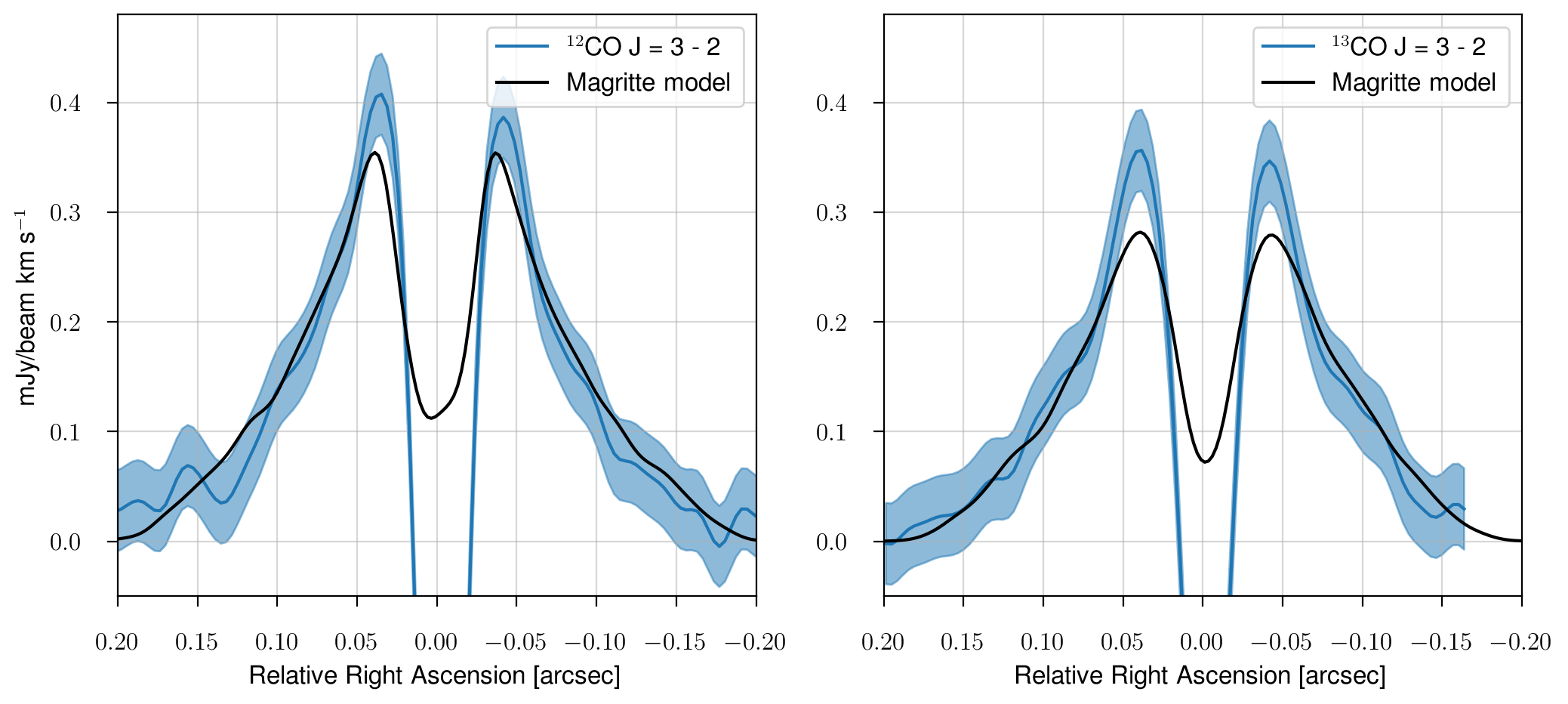}
 \caption{Slice through the moment 0 map of $^{12}$CO (left) $^{13}$CO $J = 3-2$ (right) and along the major axis of the disk (blue) together with the \textsc{Magritte} modelling results (black). 
 The range marked in blue corresponds to the rms error of the moment 0 map of the data.
  Model parameters are listed in Table \ref{table:modelparams}.}
 \label{fig:model-mom0-slice}
\end{figure*}

Together with a visual inspection of the moment 0 maps, channel maps, and line profiles, we chose the best-fitting model to the \ce{^{12}CO} moment 0 slice with a \ce{^{13}CO} abundance of $2 \times 10^{-5}$ as our final model.
Its parameters are listed in Table \ref{table:modelparams}.
Figure \ref{fig:model-mom0-slice} shows the slices through the moment 0 maps of the \ce{^{12}CO} and \ce{^{13}CO} $J = 3-2$ lines along with the \textsc{Magritte} model results.
The negative flux in at the centre of the observed lines is due to line absorption by the molecular gas located in front of the star.
The \textsc{Magritte} model's moment 0 and channel maps are shown in Appendix \ref{app:magritte}.

Fig. \ref{fig:physicalstructure} shows the density and temperature profile of the best-fitting model.
We find that \lpup's disk has a surface density profile $\Sigma(r) \sim r^{-1}$ and an $r_0$ of 6 au.
This radius corresponds to the radius where the disk's velocity profile changes from Keplerian to sub-Keplerian \citep{Kervella2016} and where the surface density steepens significantly from a power law \citep{Williams2011}.
Its midplane is warm with  $T_\mathrm{mid,1} = 900$ K, and has a shallow temperature profile characterised by $q = 0.2$.
The \ce{^{13}CO} abundance w.r.t. \ce{H2} is $2 \times 10^{-5}$, yielding a \ce{^{12}CO}/\ce{^{13}CO} ratio of 5.

\begin{table}
	\caption{The physical disk model for \lpup\ that best reproduces the data following the parametrisation of \citet{Williams2014}.
	The top parameters were fixed based on previous observations.
	The bottom parameters were obtained from modelling the $^{12}$CO and $^{13}$CO $J = 3-2$ data.
	Their ranges  are described in Sect. \ref{subsect:physics:params}.}
	\resizebox{1.0\columnwidth}{!}{%
	\centering
	\label{table:modelparams}
	\begin{tabular}{ll c} 
		\hline
    Atmospheric temperature, $T_\mathrm{atm,1}$        &   2800 K  &(1) \\
    Exponent power law $T_\mathrm{atm}(r)$, $\epsilon$ & 0.65 & (2)\\
    Inner radius of the disk, $r_\mathrm{in}$ & 2 au &(2)\\
    Velocity profile up to 6 au & Keplerian & (2) \\
    Velocity profile beyond 6 au & sub Keplerian, $\propto r^{0.853}$ & (2) \\
    Turbulent velocity & 0.5 km s$^{-1}$ & (3) \\
    \ce{^{12}CO} abundance & $1 \times 10^{-4}$ w.r.t. \ce{H2} & (3) \\
    Total gas mass of the disk, $M_\mathrm{gas}$ & $2.5 \times 10^{-4}$ M$_\odot$ & (4) \\ 
    Stellar mass, $M_*$ & $0.659\ \times$ M$_\odot$ & (2) \\
		\hline
    Midplane temperature, $T_\mathrm{mid,1}$  & 900 K&  \\
    Power-law exponent $T_\mathrm{mid}(r)$, $q$         & 0.2 &\\
    Characteristic disk radius, $r_0$ & 6 au &\\
    Power-law exponent surface density $\Sigma(r)$, $\gamma$ & -1.0& \\
    \ce{^{13}CO} abundance & $2 \times 10^{-5}$ w.r.t. \ce{H2} & \\
		\hline
	\end{tabular}
	}
    \footnotesize
    \\
    { {{References.}} (1) \citet{Danilovich2015}; (2) \citet{Kervella2016}; \\(3) \citet{Homan2017}; (4) \citet{Kervella2015}.
    }
\end{table}

\begin{figure}
\centering
 \includegraphics[width=0.9\columnwidth]{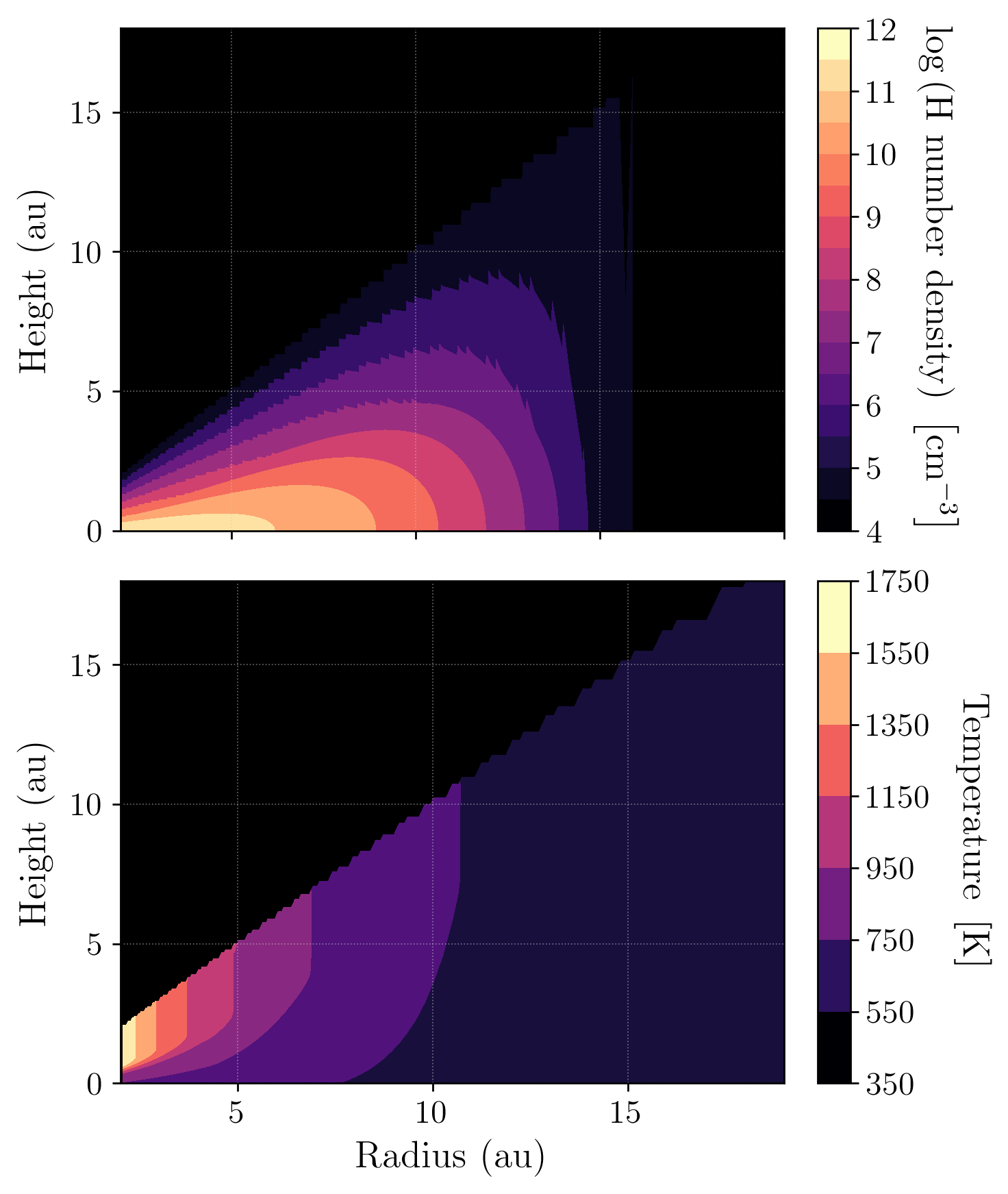}
 \caption{Density (upper panel) and temperature (lower panel) of the \textsc{Magritte} model that best reproduces the observations of \lpup's disk.  
 The parameters of the physical model are listed in Table \ref{table:modelparams}.
 }
 \label{fig:physicalstructure}
\end{figure}

\begin{figure*}
\centering
 \includegraphics[width=1\textwidth]{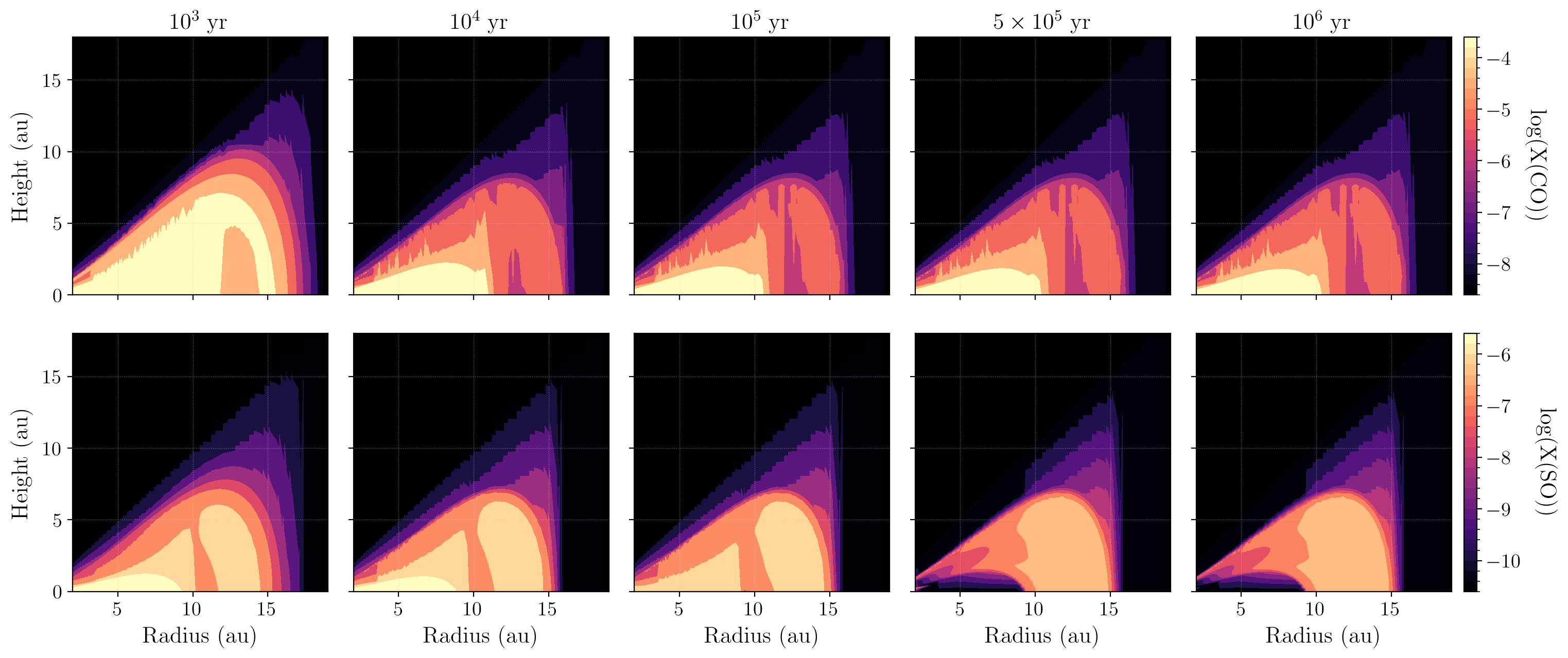}
 \caption{Fractional abundance of CO (top panels) and SO (bottom panels) throughout the disk as predicted by the chemical model.
 The colour map shows the logarithm of the abundance w.r.t. \ce{H2}.
 The different columns show the abundance maps after different times, increasing from left to right.
 }
 \label{fig:CO_SO}
\end{figure*}

\section{Chemical model}			\label{sect:chem}

The disk chemical model used is the PPD chemical model of \citet{Walsh2015}, which includes the full gas-phase chemical network of the UMIST Database for Astrochemistry (UDfA) \textsc{Rate12} \citep{McElroy2013} supplemented with a comprehensive dust-gas and grain-surface chemistry.
The \textsc{Rate12} gas-phase network includes ion-molecule and neutral-neutral reactions, photoionisation and photodissociation (parametrised to the interstellar radiation field), cosmic-ray-ionisation, and cosmic-ray-induced photoreactions.  We include all these processes by default.  
To this base network, \citet{Walsh2015} added a suite of three-body association reactions and hot neutral-neutral reactions (those with a large activation barrier) from compilations used in combustion chemistry \cite[e.g.,][]{Baulch2005}.  
Also added were a set of collisional dissociation reactions for all species expected to be abundant at the high densities and temperatures expected in inner disks (i.e., mostly saturated molecules and associated radicals). 

The protoplanetary disk chemical code also includes a treatment for self- and mutual-shielding of \ce{H2}, CO, and \ce{N2} \cite[see e.g.,][]{Heays2017}.  
To compute the shielding functions for photodissociation of \ce{H2}, CO, and \ce{N2}, we estimate the visual extinction, $A_V$ (mag), at each point in the disk by vertically integrating the \ce{H2} column from the disk surface downwards, then scaling the \ce{H2} column to $A_V$ using the standard ISM scaling ($0.5 \times (1.59 \times 10^{21}) \times A_V$; \citealt{Bohlin1978}).  
This is appropriate because the AGB star is very cool ($T_\mathrm{eff} = 2800-3500$ K, \citealt{Kervella2014,Danilovich2015}), and the only source of UV radiation is the ISM. 
We then interpolate the shielding functions provided by \citet{Heays2017} over $A_V$ and column density, assuming conservative fractional abundances for CO and \ce{N2} of $10^{-5}$ w.r.t. \ce{H2}.  
This shielding factor is then added as a pre-factor to the photodissociation rate which is calculated in the usual manner. 
The chemical model of \citet{Walsh2015} also includes a suite of X-ray reactions and grain surface processes. 
However, because there is no source of X-rays in this system, X-ray chemistry is effectively inactive  and is therefore not included.

The physical model retrieved in Sect. \ref{sect:physics} was used as input to the static chemical model. 
As there are no observational constraints, we assumed that the dust temperature is equal to the gas temperature. 
The assumption of thermally coupled dust and gas is reasonable considering the high densities.
The dust extinction at each point in the disk was calculated by combining the \ce{H2} column densities of the disk and that in the surrounding tenuous outflow in the vertical direction for each point, assuming the ISM relation between $A_V$ and \ce{H2} column density.
The chemistry is evolved for $10^6$ yr, roughly the entire AGB lifetime.

We also calculate a 1D outflow chemical model to compare the predicted disk chemistry with the chemistry in a low-density outflow. 
The smooth outflow model is the publicly available CSE model of the UDfA \textsc{Rate22} release \citep{Millar2024}\footnote{\url{https://github.com/MarieVdS/rate22_cse_code}}.
It describes a spherically symmetric outflow with a constant mass-loss rate and outflow velocity, with a power-law temperature profile. 
Following \citet{Danilovich2015}, we adopt a mass-loss rate of $\dot{M} = 1.4 \times 10^{-8}$ M$_\odot$ yr$^{-1}$, an expansion velocity of $v_\infty = 2$ km s$^{-1}$, and a stellar temperature of $T_* = 2800$ K. 
We use the same exponent for the temperature power law as in the disk physical model, $\epsilon = 0.65$.
The density and temperature throughout the outflow is shown in Fig. \ref{fig:phys_outflow}.

Table \ref{table:model-parents} lists the O-rich parent species and their initial abundances as derived from a range of observations by \citet{Agundez2020} and augmented by metal abundances \citep{VandeSande2021}.
Note that \citet{Kervella2016} found a sub-solar metallicity of $Z = 0.008$, which could reduce some of our assumed abundances by a factor $\sim 2$ \citep{Karakas2016}. 
The parent species are used as input to both chemical models.
Therefore, we assume that the material captured during the disk generation phase has the same composition as the inner wind as an approximation of the disk's initial composition. 
We evolve the chemistry for an initial $10^3$ yr as our first snapshot to allow the physical conditions to imprint on the chemistry.

The results of the chemical model are presented in the next Sections.
We first show the results for the parent species in Sect. \ref{subsect:chem:parents}, followed by the daughters in Sect. \ref{subsect:chem:daughters}.
The formation of complex organic molecules (COMs) and a C-rich reservoir in the outer regions is shown in Sects. \ref{subsect:chem:COMs} and \ref{subsect:chem:csink}, respectively.
All abundances are fractional abundances relative to \ce{H2}.

\subsection{Behaviour of the parent species}			\label{subsect:chem:parents}

Fig. \ref{fig:CO_SO} shows the abundance of the parents CO and SO throughout the disk at different time intervals. 
Their abundance distribution changes over time. 
The CO abundance decreases in the outer, more tenuous layers due to photodissociation by the interstellar UV radiation field (note that self- and mutual CO self-shielding is included, Sect. \ref{sect:chem}).
The abundance of SO decreases in the midplane up to a radius of 10 au after $5 \times 10^5$ yr, shifting its peak abundance to the outer edge of the disk. 
This is not due to depletion onto dust as the dust is too warm, but it is caused by reactions with C (liberated predominantly from CO by cosmic rays), which reacts with SO to form CS.
\ce{SO2} shows similar behaviour (Fig. \ref{fig:SO2_SiS} in Appendix \ref{sect:chem}), where reactions with C lead to the formation of SO.
The large C/S abundance ratio of $\sim 12$ makes these efficient SO and \ce{SO2} destruction pathways.
The observed emission of SO and \ce{SO2} \citep{Kervella2016} suggests that the disk is likely not older than $\sim 10^5$ yr; this is further discussed in Sect. \ref{subsect:disc:age}. 
The abundance profiles of all parents at $10^5$ yr are shown in Fig. \ref{fig:allparents}.

\begin{figure}
\centering
 \includegraphics[width=1\columnwidth]{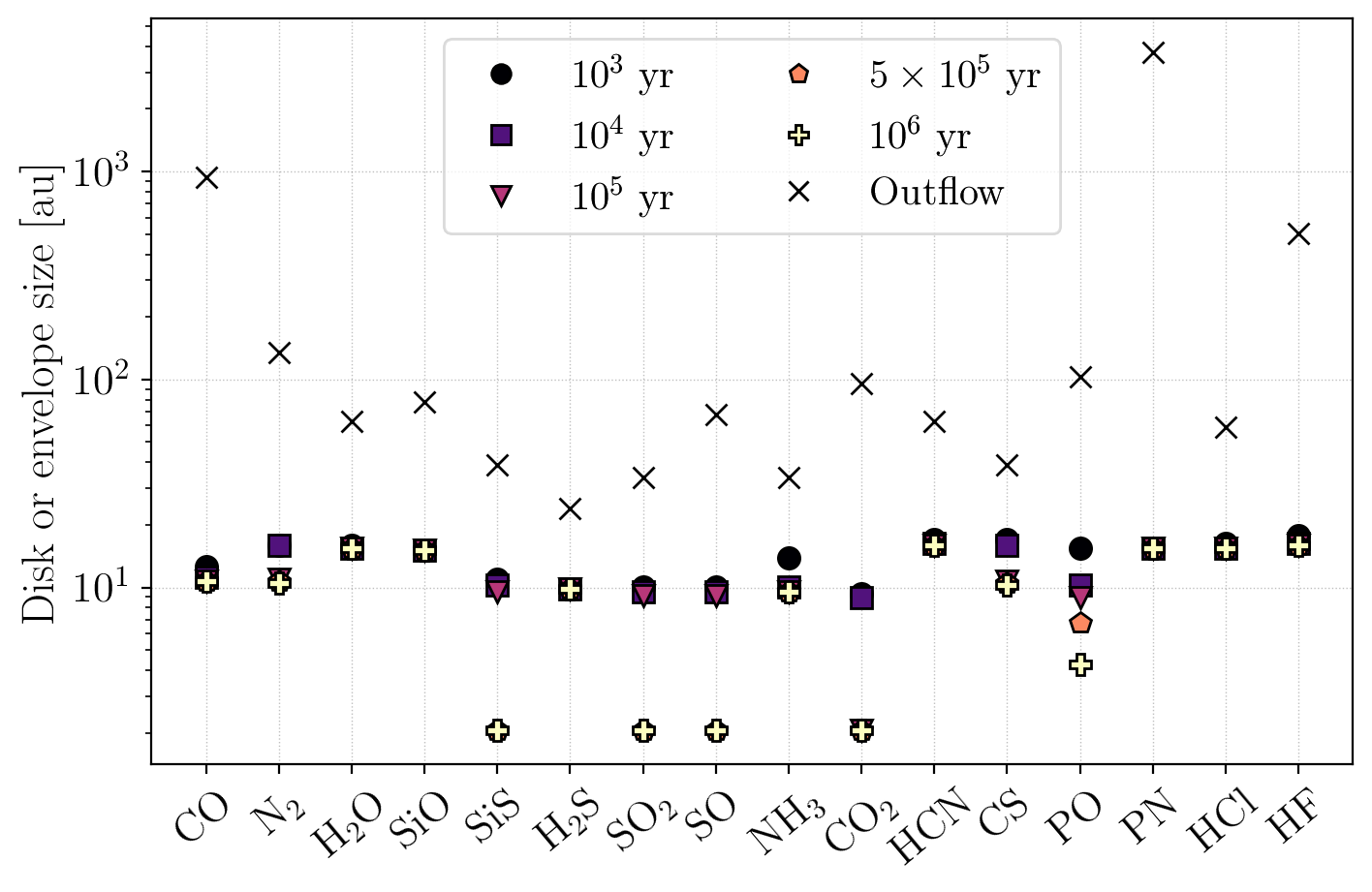}
 \caption{Radial extent of the parent species predicted by the disk and outflow chemical models in au.
 Their radial extent in the disk measured at the midplane is shown at different times, characterised by different markers and colours. 
 The grey cross shows their extent in a smooth outflow model. 
 }
 \label{fig:extent}
\end{figure}

The radial extent measured at the midplane of the parents at different times within the disk is shown in Fig. \ref{fig:extent}, together with the size of their molecular envelopes in the outflow model. 
These are the radii where the initial abundance has decreased by a factor of $e$, corresponding to the $e$-folding radius in the outflow. 
All parents have a radial extent of $\sim$ 10 au, which is consistent with the disk's density distribution (Fig. \ref{fig:physicalstructure}). 
This is smaller than the size of their molecular envelope in the outflow, which is larger by a factor of a few (e.g., for \ce{H2S} or \ce{NH3}) to two orders of magnitude (for CO and PN).
In the disk, the radial extent of most parent species does not change significantly over time.
That of the parents SiS, \ce{SO2}, SO and \ce{CO2} decreases by a factor of $\sim 4-5$ over time; that of PO decreases by a factor of $\sim 2$.
For \ce{SO2}, SO and \ce{CO2}, this is due to the abundance decreasing in the midplane followed by photodissociation in the outer layers of the disk.
The decrease in radial extent for SiS and PO is purely due to photodissociation, these species do not show a gap in the inner midplane (Fig. \ref{fig:allparents}).

\begin{figure*}
\centering
 \includegraphics[width=1\textwidth]{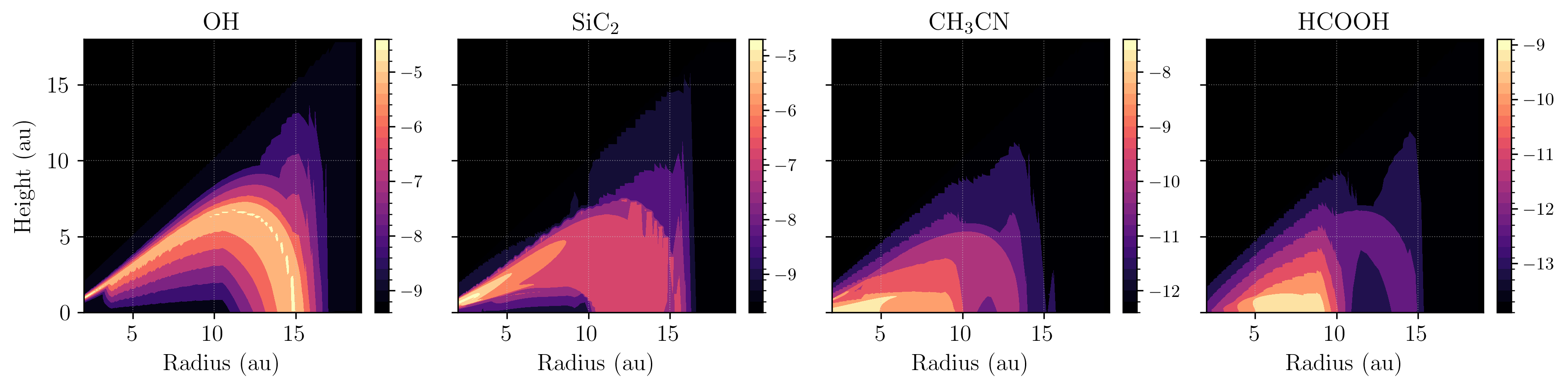}
 \caption{Fractional bundance of the daughters OH, \ce{SiC2}, \ce{CH3CN}, and HCOOH throughout the disk as predicted by the chemical model at $10^5$ yr.
 The colour map shows the logarithm of the abundance w.r.t. \ce{H2}.
 Note that each colourmap has a different dynamic range.
 }
 \label{fig:daughters}
\end{figure*}

\subsection{Behaviour of daughter species}			\label{subsect:chem:daughters}

Fig. \ref{fig:daughters} shows the abundance of the daughters OH, \ce{SiC2}, \ce{CH3CN} and HCOOH throughout the disk at $10^5$ yr.
The abundance of OH peaks in the outer regions of the disk, with abundances up to $\sim 3 \times 10^{-6}$. 
This distribution is due to its formation from the photodissociation of the parent \ce{H2O} and is similar to its behaviour in an outflow, where it forms a molecular shell.

\ce{SiC2} is also present in the outer regions of the disk, but its abundance is largest in the inner disk edge where it reaches $\sim 3 \times 10^{-6}$.
This is caused by the unique combination of high density, high temperature, and strong UV radiation field in this region.
\ce{SiC2} is formed by the reaction \ce{Si + C2H2}, where Si is produced by the photodissociation of the parent SiO.
The higher abundance in the inner disk edge is inherited from that of the reactant \ce{C2H2} (Fig. \ref{fig:daughters_ratio}) and is caused by the temperature dependencies in its formation pathway, which are elaborated on in Sect. \ref{subsubsect:disc:chem:outer}.

\ce{CH3CN} (methyl cyanide) is not located in the outer edge of the disk, but in the inner disk midplane.
Its peak abundance is $\sim 3 \times 10^{-9}$, which is two orders of magnitude larger than its predicted peak abundance in an outflow.
This C-bearing molecule is formed via \ce{CH3+} and the parent HCN producing \ce{CH3CNH+}, which then dissociatively recombines by collisions with negatively charged dust grains into \ce{CH3CN}.
\ce{CH3+} plays a crucial role in the chemistry of the inner midplane.
It is formed via \ce{CH4 + He+}, where both reactants have a cosmic ray origin, as \ce{CH4} is formed by the successive addition of hydrogen to C (liberated by cosmic rays from CO). 
Sect. \ref{subsubsect:disc:chem:midplane} elaborates on the cosmic-ray induced midplane chemistry.

The abundance of \ce{HCOOH} (formic acid) is also largest in the midplane. 
Its peak abundance is $\sim 5 \times 10^{-9}$, three orders of magnitude larger than its peak abundance in an outflow.
Its formation is also initiated by cosmic rays: \ce{H3+} (a product of cosmic-ray reactions) reacts with the parent \ce{CO} to form \ce{HCO+}.
Reaction with the parent \ce{H2O} produces \ce{HCOOCH2+} and \ce{HCOOH}.
Unlike \ce{CH3CN}, its peak abundance is not in the inner regions of the disk but between 5 and 9 au. 
This is due to the difference in destruction mechanisms: unlike \ce{CH3CN}, HCOOH is efficiently destroyed by collisional dissociation. 
The rate of this destruction mechanism increases with density, destroying HCOOH in the dense inner midplane.

\subsection{Formation of complex organic molecules}			\label{subsect:chem:COMs}

\begin{figure*}
\centering
 \includegraphics[width=0.8\textwidth]{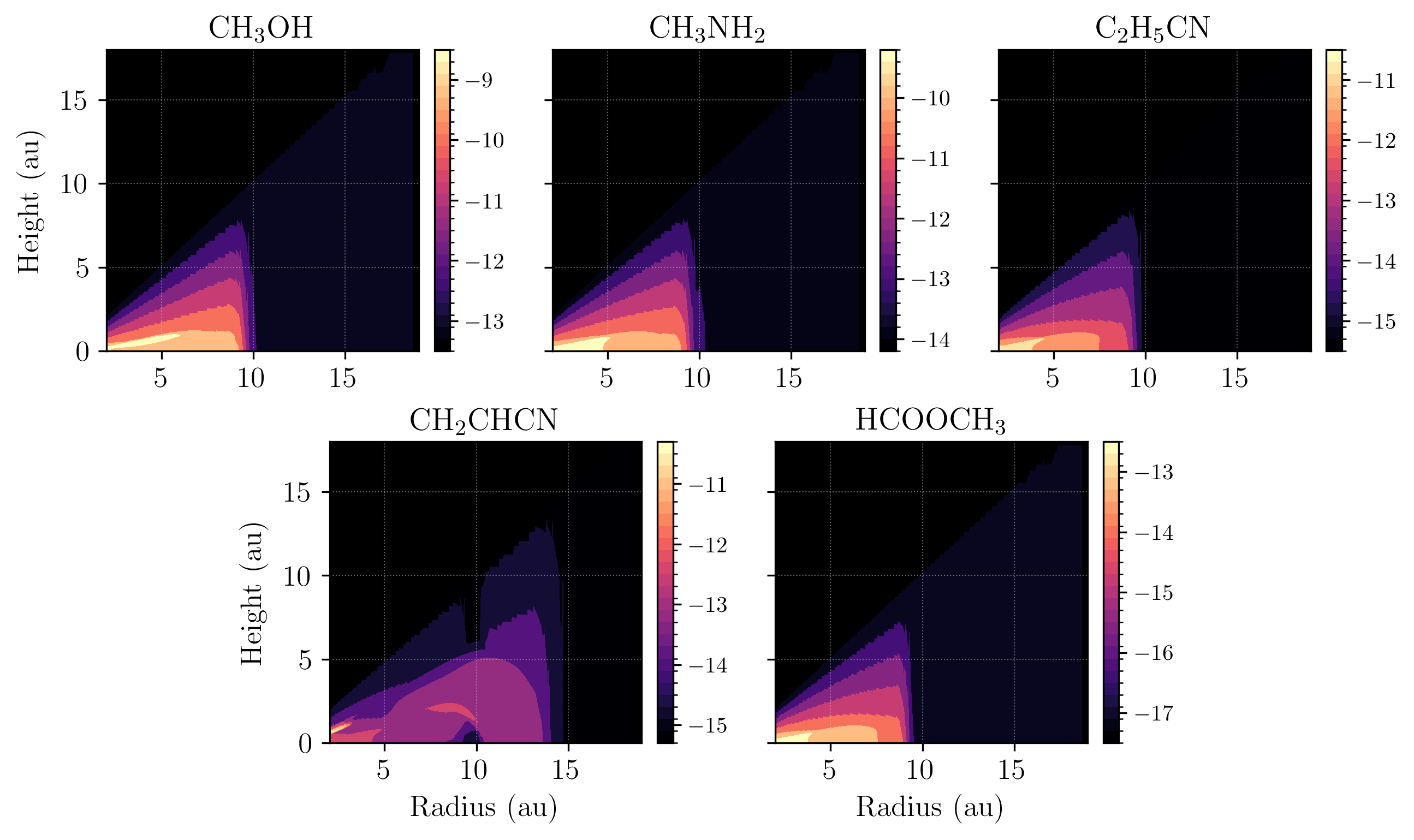}
 \caption{Fractional abundances of complex organic molecules (COMs) formed within the disk as predicted by the chemical model after $10^5$ yr.
 The colour map shows the logarithm of the abundance w.r.t. \ce{H2}. 
 Note that each colour map has a different dynamic range.
 The additional COMs \ce{CH3CN} and HCOOH are shown in Fig. \ref{fig:daughters}.
 }
 \label{fig:ab_COMs}
\end{figure*}

The COMs \ce{CH3OH} (methanol), \ce{CH3CN} (vinyl cyanide), \ce{CH3NH2} (methylamide), \ce{C2H5CN} (ethyl cyanide), \ce{CH2CHCN} (acrylonitrile), \ce{HCOOH} (formic acid), and \ce{HCOOCH3} (methyl formate) are abundantly present in the inner disk and are formed by cosmic-ray induced chemistry in the midplane (Sect. \ref{subsubsect:disc:chem:midplane}). 
Their abundance distribution at $10^5$ yr is shown in Fig. \ref{fig:ab_COMs}.

\begin{figure}
\centering
 \includegraphics[width=0.8\columnwidth]{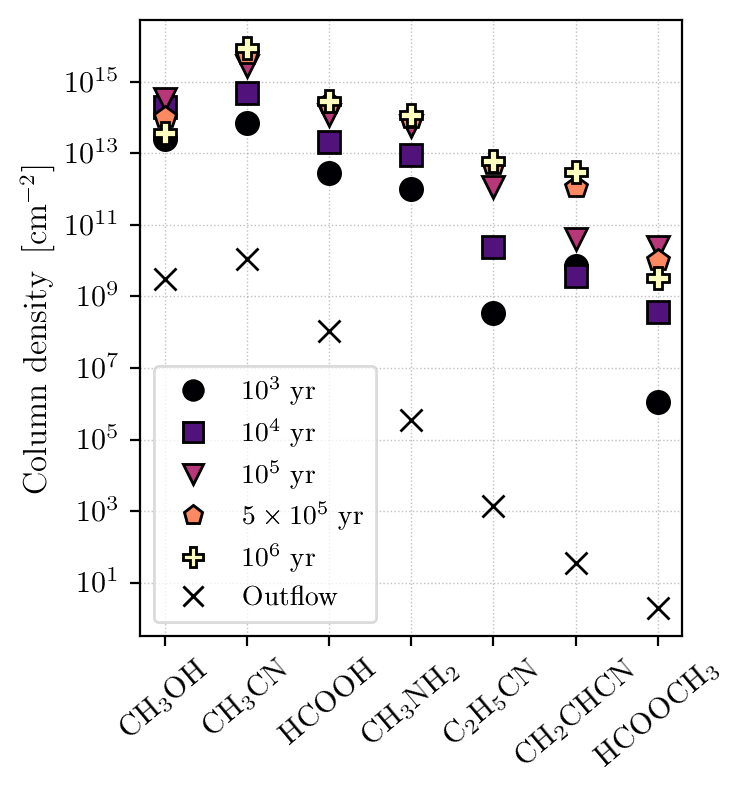}
 \caption{Disk-averaged column densities of COMs produced in the chemical model of the disk at different times, characterised by different markers and colours.
 The grey cross shows the results from a smooth outflow model. 
 }
 \label{fig:CD_COMs}
\end{figure}

Fig. \ref{fig:CD_COMs} shows their column density at different times in the disk model, together with their column density in the outflow model.
The column density of species in the disk is a disk-averaged column density, calculated via
\begin{equation}	\label{eq:CD}
	\bar{N} = \frac{2}{\pi r_D^2} \times \sum_{i=0}^{D} 2 \pi r_i \Delta r_i N_i,
\end{equation}
where $N_i$ is the vertical column density within $\Delta r_i$ and the factor 2 accounts for the symmetry of the model across the midplane. 
At $10^5$ yr, all COMs listed above have a disk-averaged column density larger than $10^{10}$ cm$^{-2}$. 
\ce{CH3CN} has the largest column density of $1.80 \times 10^{15}$ cm$^{-2}$, followed by \ce{CH3OH} with $3.37 \times 10^{14}$ cm$^{-2}$ and \ce{HCOOH} with $1.19 \times 10^{14}$ cm$^{-2}$. 
The column densities are at least five orders of magnitude larger than those predicted by the outflow model.
The largest differences are seen for \ce{CH3NH2}, \ce{C2H5CN}, \ce{CH2CHCN}, and \ce{HCOOCH3}, with a difference of at least 8 orders of magnitude, showing that the chemistry in the disk is favouring the formation of larger saturated complex molecules.

\subsection{Carbon-rich reservoir in the outer regions}			\label{subsect:chem:csink}

Carbon-bearing molecules are abundantly formed in the outer regions of the disk.
CO and other C-bearing parents are photodissociated by interstellar UV photons, liberating the reactive C atom.
Carbon-chains, e.g., \ce{C2H2} and \ce{HC3N}, are abundantly formed in this region (Fig. \ref{fig:daughters_ratio}).
Longer highly-unsaturated carbon-chains, such as \ce{C10H2}, \ce{C8H}, and \ce{HC9N}, are also produced.
However, the chemical reaction network is geared towards interstellar chemistry and does not include full saturation of these carbon-chains. 
Hence, the predicted abundances of these longer chains are likely over-estimated.

\begin{figure*}
\centering
 \includegraphics[width=1.0\textwidth]{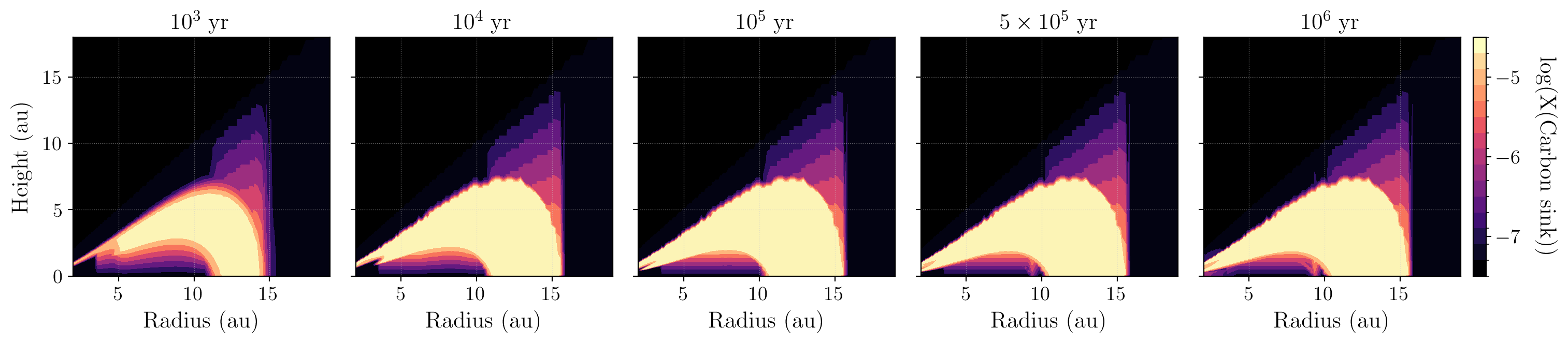}
 \caption{Total fractional abundance of all species with 6 or more C atoms (called carbon sinks, Sect. \ref{subsect:chem:csink}) in the disk as predicted by the chemical model. 
 The colour map shows the logarithm of the abundance w.r.t. \ce{H2}. 
 The different columns show the abundance maps after different times, increasing from left to right.
 }
 \label{fig:Csink}
\end{figure*}

We therefore grouped all species with 6 or more C atoms into a ``carbon sink'' reservoir. 
Fig. \ref{fig:Csink} shows the abundance of this sink particle throughout the disk at different times in its evolution. 
The total abundance of carbon-bearing species is large in the outer disk, with a maximum abundance of $\sim 3 \times 10^{-5}$.
This is remarkably large considering the disk is O-rich.
Unlike in the outer regions of an AGB outflow, newly liberated C can readily react and form (long) carbon chains thanks to the higher densities in the disk's outer region.

\section{Discussion}			\label{sect:disc}

The physical structure of the disk has a large influence on the chemistry around \lpup.
In Sect. \ref{subsect:disc:phys}, we compare our newly derived physical structure to PPDs and disks around post-AGB stars. 
We then compare our results to those of \citet{Homan2017} and suggest future observations to better constrain the density and temperature within the disk. 
This is important, as the predicted chemistry depends on the physical structure.
In Sect. \ref{subsect:disc:compobs}, we qualitatively compare our predicted abundance profiles to the previous observations of \citet{Kervella2016} as a first-order test of the model.
We elaborate on using the predicted SO and \ce{SO2} abundance profiles to age the disk at $\sim 10^5$ yr in Sect. \ref{subsect:disc:age}.
Finally, the chemistry within the disk is discussed in Sect. \ref{subsect:disc:chem}, where we examine the surprisingly rich C-rich chemistry in the O-rich disk and suggest molecular tracers of the disk structure.

\subsection{Physical structure of the disk}			\label{subsect:disc:phys}

We compare our retrieved physical structure of the disk around \lpup\ to PPDs and post-AGB disks in Sect. \ref{subsubsect:disc:phys:ppd}.
A comparison to the previous model of \citet{Homan2017} is made in Sect. \ref{subsubsect:disc:phys:homan}.
Since our physical model was retrieved using just two molecular emission lines, we describe the observations needed for a further refinement of the model in Sect. \ref{subsubsect:disc:phys:obs}.

\subsubsection{Comparison to PPDs and post-AGB disks}			\label{subsubsect:disc:phys:ppd}

We retrieved that the disk around \lpup\ has a surface density profile $\Sigma(r) \sim r^{-1}$ and is warm, with a midplane temperature of 900 K and $T_\mathrm{mid}(r) \sim r^{-0.2}$.
The exponent of the surface density distribution, $\gamma$, is on the shallower end of those found for PPDs, for which $\gamma \lesssim -1$ \citep{Andrews2020}.
Disks around post-AGB stars have $-3 \leq \gamma \leq -1$, although it is difficult to constrain this value with infrared interferometry alone \citep{Hillen2014,Hillen2015,Corporaal2023}.
Our value of $\gamma = -1$ hence also agrees with post-AGB disks and appears to suggest that the density profile of this disk around an AGB star lies between that of a PPD and post-AGB disk.

%
%

The midplane temperature of \lpup's disk is much warmer than in PPDs: $T_\mathrm{mid,1} = 900$ K compared to $100-300$ K for PPDs. 
The power-law is also flatter, with $q = 0.2$ rather than $q = 0.5 - 0.7$ for PPDs \cite[e.g.,][]{Andrews2009,Williams2014}.
The temperatures of post-AGB disks lie within $1400 - 3600$ K, which are single blackbody temperatures for the disk as it is, again, difficult to constrain their midplane temperatures using infrared interferometry alone \citep{Kluska2018,Corporaal2023}. 
The disk around \lpup\ again lies between these two ranges: warmer than PPDs, colder than post-AGB disks.
This is expected as the gas ejected by the AGB star is warm and its high luminosity can provide a lot of dust heating.
In post-AGB stars, the stripped star is much warmer, leading to larger temperatures.

\subsubsection{Comparison to Homan et al. (2017)}			\label{subsubsect:disc:phys:homan}

We  used a parametric model commonly used for PPDs to model the same observations used by \citet{Homan2017} as we were unable to reproduce their model from the information provided in that work.
We did use the same velocity profile of \citet{Kervella2016} and also adopted certain observational parameters of \citet{Homan2017}, as described in Sect. \ref{subsect:physics:params}.

We find three main differences between our model and theirs:
\begin{enumerate}
	\item \citet{Homan2017} assumed a three-staged radial temperature profile, which includes an exponential and arctan. 
This profile includes a feature the authors attribute to the heating of the gas by the dust disk.
The parameter space explored to achieve this fit is not explained.
We find that a power-law expression (Eq. \ref{eq:Tatm}) is sufficient to reproduce the observations, indicating that their analytical expression could be overfitted and may have been more complex than needed.
	\item \citet{Homan2017} retrieved a steep radial density profile at the midplane of $\rho(r,z=0) \sim r^{-3.1}$, which is much steeper than our $\rho(r,z=0) \sim r^{-0.4}$ (Eqs \ref{eq:sigma}, \ref{eq:rho}, \ref{eq:H}).
The steepness of their density profile does not agree with observations of either PPDs or post-AGB disks. 
The large difference in density profile could be due to the differences in temperature structures.
	\item The \ce{^{13}CO} abundance retrieved by \citet{Homan2017} is $1 \times 10^{-5}$, we find a value of $2 \times 10^{-5}$ w.r.t. \ce{H2}. 
We used the same \ce{^{12}CO} abundance of $1 \times 10^{-4}$ w.r.t. \ce{H2}. 
This leads to different \ce{^{12}CO}/\ce{^{13}CO} ratios: 10 by \citet{Homan2017}, 5 in this work. 
While both values lie within the range found by \citet{Ramstedt2014}, our value of 5 is on the low side.
Optical depth effects might cause such a low \ce{^{12}CO}/\ce{^{13}CO} ratio.
Note that photodissociation effects due to differences in self-shielding between the isotopologues would increase the ratio \citep{Saberi2019}.
Additionally, a factor of 2 difference is reasonable within the uncertainties of modelling \citep{Ramstedt2008}.  
\end{enumerate}

\subsubsection{Refining the physical structure: further observations}			\label{subsubsect:disc:phys:obs}

Besides \ce{CO} and \ce{^{13}CO}, the available ALMA data includes rotational transitions of \ce{H2O}, \ce{^{29}SiO}, \ce{^{30}SiO}, SiS, SO, and \ce{SO2}.
The isotopologues of SiO were used to constrain the velocity structure of the disk by \citet{Kervella2016}.
While the other lines shed some light on the chemical structure of the disk (Sects. \ref{subsect:disc:compobs} and \ref{subsect:disc:age}), further observations are necessary to better constrain the disk's physical structure, given the differences with \citet{Homan2017} (Sect. \ref{subsubsect:disc:phys:homan}).
A better constraint of the physical structure of \lpup's disk will address the differences with the \citet{Homan2017} model and is a crucial input to the chemical model and its predictions.

Spatially resolved emission of abundant optically thin CO isotopologues, such as \ce{C^{18}O}, \ce{C^{17}O}, and  \ce{^{13}CO} \citep{Danilovich2017H2O,DeNutte2017}, can help determine the disk gas mass and its density distribution more accurately as they probe different layers within the disk.
Similarly, \ce{H2CO} \cite[e.g.,][]{Qi2013,Pegues2020}, \ce{HCN} \cite[e.g.,][]{Long2021}, \ce{HC3N}, \ce{CH3CN}, and \ce{C3H2} \cite[e.g.,][]{Bergner2018,Ilee2021}, are well-known temperature probes in PPDs. 
Our chemical model predicts column densities of $3.72 \times 10^{14}$, $8.54 \times 10^{17}$, $4.21 \times 10^{13}$, $2.80 \times 10^{15}$, and $3.65 \times 10^{12}$ cm$^{-2}$, respectively, for these molecules.

\ce{CH3CN} and \ce{C3H2} have yet not been detected towards O-rich AGB stars: their detection would put strong constraints on our chemical model.
\ce{HC3N} has only been detected around OH 231.8+4.2, which has a bipolar outflow with velocities up to $\sim 400$ km s$^{-1}$ \citep{VelillaPrieto2015}.
So far, \ce{H2CO} has only been detected around the moderate mass-loss rate O-rich AGB star IK Tau with a column density of $2.8 \times 10^{14}$ cm$^{-2}$, which is close to our predicted column density \citep{VelillaPrieto2017}.
IK Tau likely has a stellar companion shaping its outflow \citep{Coenegrachts2023}.
\ce{H2CO} has also been detected in the Rotten Egg Nebula, an AGB star with an A-type companion \citep{Lindqvist1992}, and the Helix Nebula, which might host an unseen dwarf M star \citep{Gruendl2001,Tenenbaum2009,Zack2013}. 
It seems that detections of \ce{H2CO} around evolved stars appear to suggest deviations from spherical symmetry.

\subsection{Comparison of the chemical model to observations}			\label{subsect:disc:compobs}

We qualitatively compare our chemical modelling results with the observations of \citet{Kervella2016}.
The observed \ce{^{12}CO} emission (Appendix A in \citealt{Kervella2016}) has a radial extent of  $\sim$ 9 au and a vertical extent of $\sim$ 3 au.
The peak CO abundance predicted by our chemical model lies within $\sim$ 11 au radially and about 2.5 au vertically (Fig. \ref{fig:CO_SO}). 
Keeping in mind that we are comparing abundances with observations, the CO abundance predicted by the chemical model is consistent with the observed \ce{^{12}CO} emission.
A standard set of parent species and their abundances was used as input when running the chemical model. 
The initial CO abundance in the chemical model is a factor 3 larger than that assumed in the physical model, following \citet{Homan2017} ($3 \times 10^{-4}$ versus $1 \times 10^{-4}$ w.r.t. \ce{H2}, Sect. \ref{sect:chem}). 
As we did not a priori know how the chemical abundances would change over time, nor which abundance (profile) to assume when retrieving the physical structure from the observations, we kept these values when calculating the physical and chemical structures.
We find that the results of the chemical model are consistent with its independently derived physical structure.
The disk averaged CO abundance at $10^5$ yr decreased to $2.86 \times 10^{-4}$ w.r.t. \ce{H2}.

Besides the \ce{^{12}CO} and \ce{^{13}CO} $J = 3-2$ lines, \cite{Kervella2016} also presented the \ce{SO2} ($v = 0, J_{K_a,K_c} = 34_{3,31} - 34_{2,32}$), SO ($\nu = 0, N_J = 8_8 - 7_7$), and SiS ($\nu = 1, {J} = 19-18$) lines (their Appendices C - E).
The \ce{SO2} and SO emission is more extended vertically (about 6 au and $\sim$ 3 au, respectively) compared to the vertically compact SiS emission (up to 2 au).
The \ce{SO} abundance of the chemical model is shown in Fig. \ref{fig:CO_SO}, that of \ce{SO2} and SiS in Fig. \ref{fig:SO2_SiS}.
At $10^5$ yr, the region of highest \ce{SO2} abundance has a vertical extent of $\sim$ 6.5 au, that of SO about $\sim$ 5.5 au and of SiS about $\sim$ 1.5 au.
The chemical model hence also qualitatively reproduces the relative vertical extents of SO, \ce{SO2}, and SiS, though a full radiative transfer model is required to check. 

Methanol has been detected towards the post-AGB star HD 101584 \citep{Olofsson2017} and OH 231.8+4.2, a bipolar nebula around an AGB star \citep{SanchezContreras2018}.
It is detected in the high-velocity outflows of these objects and is thought to originate from post-shock chemistry, where shocks in the outflow desorb the molecule from an icy grain mantle. 
The retrieved column densities of \ce{CH3OH} are $\sim 10^{16}$ and $1-8 \times 10^{14}$ cm$^{-2}$, corresponding to fractional abundances of $\sim 3 \times 10^{-6}$ and $10^{-8} - 10^{-7}$ w.r.t. \ce{H2}, respectively.
Our chemical model predicts a column density of $\sim 10^{14}$ cm$^{-2}$ around $10^5$ yr, with a peak fractional abundance around $3 \times 10^{-9}$ w.r.t. \ce{H2}.
The strongest \ce{CH3OH} line detected in HD 101584, the $8_{-1}$ - $7_0$ line at 229.759 GHz, is covered by the 2016.1.00207.S. dataset (PI P. Kervella), with a peak flux around 80 mJy. 
It is not clearly detected in the archival \lpup\ data, with an upper limit for the peak flux of around $2-3$ mJy. 
From running a simplified radiative transfer calculation, using the accelerated lambda iteration method code (ALI, e.g., \citealt{Danilovich2021}, with molecular data from \citealt{Rabli2010}), based on the spherically symmetric model of \citet{Danilovich2015} and assuming a Gaussian abundance profile with an $e$-folding radius of $10^{15}$ cm, we determine that for a methanol abundance of $3 \times 10^{-9}$ w.r.t. \ce{H2} (the maximum abundance predicted by the chemical model) the $8_{-1}$ - $7_0$ line would not be detected above the noise for \lpup.
The non-detection of \ce{CH3OH} in \lpup\ could be due to its lower abundance or due to a different excitation mechanism, as methanol is formed in the warm inner midplane of the disk via gas-phase reactions rather than via shock chemistry in a high-velocity outflow, as in the two other objects.
Further observations are necessary to confirm the (non)detection of methanol in \lpup's disk.

A radiative transfer analysis of the chemical model predictions lies outside the scope of this paper. 
Moreover, it is important to note that the predicted chemistry depends on the input physical structure. 
A parameter study to quantify the impact of different physical structures on the chemistry is computationally intensive and will be addressed in future work.
As stated in Sect. \ref{subsubsect:disc:phys:obs}, further observations are necessary to constrain the retrieved physical model of \lpup's disk.

\begin{figure*}
\centering
 \includegraphics[width=1\textwidth]{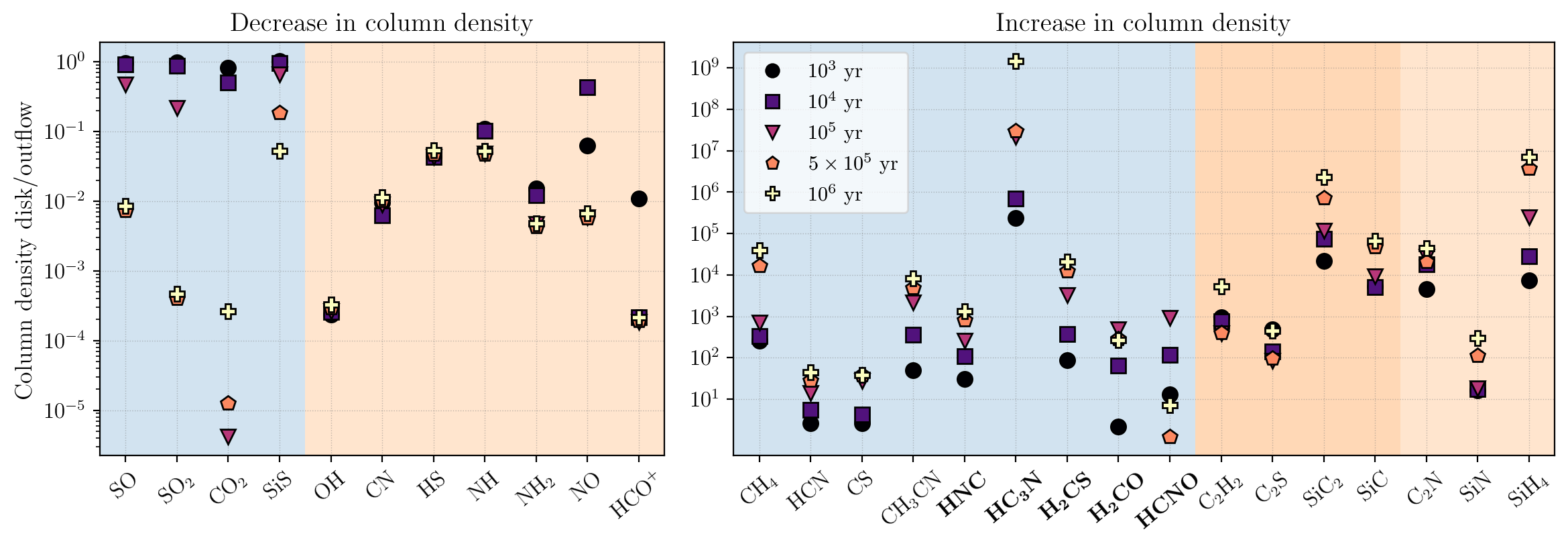}
 \caption{Ratio of the disk column density relative to CO over the outflow column density relative to CO (Eq. \ref{eq:ratio}) for species that show a decrease (left panel) and increase in column density relative to the outflow (right panel) of more than an order of magnitude at any time in the disk chemical model.
 Different markers and colours show the ratio at different times for the disk model.
 The shaded regions group species originating from different types of chemistry.
 Blue: midplane (cosmic ray) chemistry, orange: outer edge (photo) chemistry, dark orange: hot inner edge.
 Boldfaced labels denote species that are formed via cosmic ray and photochemistry.
 These species can be regarded as chemical tracers of the disk.
  }
 \label{fig:coldens-ratio}
\end{figure*}

\subsection{Age of the disk}			\label{subsect:disc:age}

After $5 \times 10^5$ yr, the chemical model predicts a gap in the inner midplane for SO and \ce{SO2} (Figs. \ref{fig:CO_SO} and \ref{fig:SO2_SiS}), which is about 1 au in height and 9 au radially for both molecules.
Such a gap is not seen in the emission maps of \ce{SO2} ($v = 0, J_{K_a,K_c} = 34_{3,31} - 34_{2,32}$) and SO ($\nu = 0, N_J = 8_8 - 7_7$, Appendices C and D in \citealt{Kervella2016}).
The beam size of the observations is $17.7 \times 14.5$ mas $\approx 1.1 \times 0.9$ au. 
The gap predicted by the chemical models should hence be observable, especially as the upper energy level of the \ce{SO2} transition is 582 K.
Therefore, under the assumptions of our chemical model of a disk that is formed at a specific time and then evolves in isolation of the outflow, we expect it to be not older than $10^5$ yr.
Hydrodynamical simulations suggest that while the disk is formed on a timescale of a few 100 yr, some outflowing gas will still be captured by and incorporated into the disk \citep{Chen2016}.
If the disk is continually refreshed with new material, its age could be older.

\lpup\ is currently in the thermally pulsing-AGB (TP-AGB) phase \citep{Kervella2016}, which is a short-lived phase of about $0.5 \times 10^5$ yr \citep{Rosenfield2014}. 
Assuming a static model, the disk could therefore both be a remnant from the early-AGB phase or have originated in the TP-AGB phase.
The timeframe of $10^5$ yr is too short for the massive planet to have been formed within this disk (which would take $\sim$10 Myr if formed via core accretion, \citealt{Pollack1996}), and is likely to be a first-generation planet present before the AGB phase.

\subsection{Chemistry within the AGB disk}			\label{subsect:disc:chem}

The chemistry within the disk around \lpup\ is different to that in a low-density O-rich outflow both in distribution and composition.
The only similarity is the lack of dust-gas chemistry, although this is due to different factors.
In the outflow, the density is too low for gas-phase species to efficiently accrete onto the dust.
The density in the disk midplane is at least four orders of magnitude larger, but the dust temperature is too high for accreted ices to remain on the dust surface.
The dust temperature was taken to be that of the gas, a reasonable assumption given the density of the disk.
However, a lower dust temperature could result in the depletion of gas-phase species and the formation of ices.

To compare the chemical composition of the disk to that of the outflow, we use the ratio of their column densities.
This column density ratio is weighted by their respective CO column densities,
\begin{equation}	\label{eq:ratio}
	\mathrm{Ratio} = \frac{\bar{N}_X}{\bar{N}_{\mathrm{CO}}} \times \frac{N_\mathrm{CO}}{N_X},
\end{equation}
with $\bar{N}_X$ the disk-averaged column density of species X (Eq. \ref{eq:CD}) and $N_X$ the column density of species $X$ in the outflow.
This normalises the ratio to the most abundant molecule (after \ce{H2}), allowing for a better comparison between the chemistries of these different density distributions.
The CO column density ratio, $\bar{N}_\mathrm{CO}/N_\mathrm{CO}$, is $\sim 100$ for all times in the disk model, decreasing from 126 at $10^3$ yr to 98 at 10$^6$ yr.

The chemical composition of the disk is different from that of the outflow: the disk does not simply inherit the initial composition, but also significantly modifies the abundances of parent and daughter species.
Unlike in the outflow and despite its O-rich composition, numerous C-bearing molecules are abundantly formed throughout the disk, including COMs. 
Two main chemical regions can be distinguished: (i) the dense midplane, where chemistry is initiated by cosmic rays, and (ii) the outer edge, where it is initiated by interstellar UV photons.
Within the outer region, the hot inner edge displays a different chemistry than the cooler outer regions.
The chemistry in the dense midplane differs from that in a PPD because of the higher temperatures in the disk, while the chemistry of the outer regions is different to an AGB outflow due to the higher densities in this region.

Fig. \ref{fig:coldens-ratio} shows the column density ratio for a set of species which have a column density larger than $10^{10}$ cm$^{-2}$ at some time in the disk and which show a difference of at least an order of magnitude at some time in the disk chemical model.
Their abundance profiles are shown in Fig. \ref{fig:daughters_ratio} and the absolute values of their column densities are shown in Fig. \ref{fig:coldens-value}.
These species can be regarded as chemical tracers of the presence of a disk structure.
The left panel shows species which have a smaller column density in the disk model compared to the outflow model, the right panel shows those which have a larger column density.
The shaded regions group species originating from different types of chemistry: midplane (cosmic ray) chemistry is highlighted in blue, outer edge (photo) chemistry in orange, with species present in the hot inner edge highlighted in dark orange.
Species formed via both cosmic ray and photochemistry are placed in the blue region with a boldface label.
Sects. \ref{subsubsect:disc:chem:midplane} and \ref{subsubsect:disc:chem:outer} elaborate on the cosmic-ray induced chemistry in the midplane and in the photochemistry in outer regions, respectively.


\subsubsection{Cosmic-ray induced chemistry}			\label{subsubsect:disc:chem:midplane}

Cosmic rays dissociate parent species in the midplane, liberating atoms which participate in a rich chemistry thanks to the high densities in this region.
A large role is played by the reactive C, produced mostly by the destruction of the parent CO.
Species whose column density is significantly impacted by this chemistry are shown in the blue shaded regions of Fig. \ref{fig:coldens-ratio}.

\paragraph*{Decrease in column density relative to the outflow}

Reactions between C and the parents SO and \ce{SO2} form CS, leading to a gap in abundance in the dense midplane and a lower column density relative to the outflow.
At $10^5$ yr, their column densities are $\sim 5$ and $\sim 9$ times smaller than in the outflow, increasing to 100 times and $\sim 7 \times 10^3$, respectively.
\ce{CO2} shows a similar gap in the midplane, which is caused by reactions with Si forming SiO. 
Its column density is more than 5 orders of magnitude smaller than in the outflow at $10^5$ yr.
At longer timescales, this difference is reduced to roughly 3 orders of magnitude thanks to its reformation by collisional dissociation of HCOOH (see below).

The decrease in column density of SiS, reaching about an order of magnitude after $5 \times 10^5$ yr, is caused by reactions with \ce{H3+}, a product of cosmic-ray interactions. 
The resulting \ce{HSiS+} reacts with the parent \ce{H2O}, producing \ce{SiOH+} which goes on to react with the parent \ce{NH3}, funnelling Si away from SiS into SiO.

\paragraph*{Increase in column density relative to the outflow}

Hydrogen is efficiently added to carbon at the high densities and temperatures in the midplane, forming \ce{CH4}.
Its column density is about 3 orders of magnitude larger in the disk.
\ce{CH4} reacts with \ce{N+}, formed via cosmic rays, producing \ce{HCNH+}. 
This molecule then reacts with \ce{H2S} and \ce{NH3} to form HCN and HNC, with column densities more than 1 and 3 orders of magnitude larger than in the outflow, respectively.
The CS abundance increases thanks to the reactions of C with \ce{SO2} and SO, increasing its column density by about 1 order of magnitude.

\ce{HCO+} is produced by cosmic ray chemistry via \ce{H3+ + CO}. 
Radiative association of \ce{HCO+} with \ce{H2O}  yields \ce{HCOOH2+}, which reacts with \ce{H2O} and \ce{NH3} to form HCOOH (increase in column density $\gtrsim$ 4 orders of magnitude compared to the outflow).

The \ce{CH3+} cation plays a crucial role in the chemistry of the midplane.
It is formed via \ce{CH4 + He+}, both products of cosmic rays, forming \ce{CH2+} which reacts with \ce{H2} to \ce{CH3+}.
The formation of all detected COMs (and \ce{HC3N}) is linked to this cation.
The high densities of the midplane allow for the slow radiative association reactions, dissociative recombination with dust, and collisional dissociation reactions in these pathways to be efficient.
The main formation pathways of COMs via reactions with \ce{CH3+} are:
\begin{itemize}
	\item Radiative association between the parent HCN and \ce{CH3+} yields \ce{CH3CNH+}, which dissociatively recombines with negatively charged dust grains to form \ce{CH3CN} (increase in column density of $\gtrsim$ 3 orders of magnitude at $10^5$ yr compared to the outflow).
	\item \ce{NH3} can radiatively associate with \ce{CH3+} forming \ce{CH3NH3+}, which dissociatively recombines with dust to \ce{CH3NH2} (column density $\gtrsim$ 6 orders of magnitude).
The association with \ce{CH3+} also forms \ce{C2H5CNH+}, which dissociatively recombines with dust to \ce{C2H5CN} ($\gtrsim$ 6 orders of magnitude). 
A sequence of collisional dissociations and reactions with C lead to the formation of \ce{CH2CCH}, which reacts with N to form \ce{HC3N} ($\gtrsim$ 7 orders of magnitude).
	\item Reactions of \ce{CH3+} with the parent \ce{H2O} gives \ce{CH3OH2+}, which produces \ce{CH3OH} by reacting with the parent \ce{NH3} ($\gtrsim$ 3 orders of magnitude).
Subsequent reaction with \ce{H2CO} results in \ce{H5C2O2+}, which dissociatively recombines with dust to \ce{HCOOCH3} ($\gtrsim$ 8 orders of magnitude).
	\item The formation of methanol initiates the production of \ce{H2CO}, increasing its column density by almost 3 orders of magnitude at $10^5$ yr, which is formed by collisional dissociation of \ce{CH2OH} and \ce{CH3O}.
Both have their origin in the \ce{CH3+ + H2O} reaction, which forms both \ce{CH3OH} and \ce{CH3OH2+}.
\ce{CH2OH} is formed via \ce{CH3OH + H}, \ce{CH3O} is formed after \ce{CH3OH2+} reacting with \ce{CH3OH} followed by dissociative recombination with dust and collisional dissociation.
\end{itemize}

Other species formed by the successive addition of hydrogen to carbon also play an active role in the cosmic-ray induced midplane chemistry.
The column density of \ce{H2CS} increases by more than 3 orders of magnitude at $10^5$ yr. 
It is formed via \ce{CH4 + S+} (both with a cosmic-ray origin) producing \ce{H3CS+}, which then dissociatively recombines with dust.
HNCO is formed by \ce{CH2 + NO} and \ce{C2H5CN + OH}, with a increased column density of about 3 orders of magnitude at $10^5$ yr. 
The formation of \ce{HC3N} is kickstarted by \ce{CH4 + CH3+}, forming \ce{C2H5+}, which yields \ce{C2H4} after reaction with the parents \ce{H2O} and \ce{H2S}. Subsequent reactions with C and N finally produce \ce{HC3N}.


\subsubsection{Interstellar UV photochemistry}			\label{subsubsect:disc:chem:outer}

The chemistry in the more tenuous outer regions of the disk is a rich photochemistry initiated by interstellar UV photons. 
However, because of the higher densities and temperatures of the disk, the products of this photochemistry are different to those of the photochemistry taking place in the outer regions of the outflow.
\ce{C2H2} plays an important role in the chemistry of the outer regions of the disk.
It is formed by the successive addition of hydrogen to C (produced by photodissociation of mainly the parent CO) into CH, which then reacts with C again to form \ce{C2}.
These reactions lead to the formation of \ce{C2H} and finally \ce{C2H2}.
Species whose column density is significantly impacted by this chemistry are shown in the orange shaded regions of Fig. \ref{fig:coldens-ratio}.

\paragraph*{Decrease in column density relative to the outflow}

The daughters OH, CN, HS, NH, and \ce{NH2} are all first-generation daughters formed by photodissociation of the parents \ce{H2O}, HCN, \ce{H2S}, and \ce{NH3}. 
They are present in a shell in the outer disk. 
Their main destruction mechanisms are photodissociation and reaction with \ce{H2} to reform the parent species in a photodissociation - hydrogen addition loop, depending on the density.
The significant decreases in their column densities relative to the outflow is caused by reactions funnelling them away from their parents.

Reactions with C-bearing species remove OH and CN from their photodissociation - hydrogen addition loops, reducing their column densities by $\sim 5 \times 10^3$ and 2 orders of magnitude, respectively.
OH mainly reacts with C to reform CO and with \ce{CH2} to produce \ce{H2CO}, which is also formed by \ce{O + CH3}.
CN is involved in reactions with \ce{C2H2} and \ce{C2H4}, producing cyanopolyynes and CN-bearing C-chains.

The column densities of HS, NH and \ce{NH2} decrease between one and two orders of magnitude relative to the outflow.
S is removed from the photodissociation - hydrogen addition loop by the formation of CS via \ce{HS + C} and S + \ce{CH2}, and by the formation of \ce{H2CS} via S + \ce{CH3}.
NH reacts with O to produce NO, which is then efficiently destroyed by Si forming SiO and NH forming \ce{N2O}, which photodissociates into \ce{N2}.
Subsequently, the column density of NO is about two orders lower in the disk after $10^5$ yr.
NH also efficiently reacts with S to produce NS, which then reacts with O to form NO and SO. 

The column density of the cation \ce{HCO+} is almost 4 orders of magnitude smaller in the disk.
Its main formation mechanism is \ce{C+ + H2O}. 
However, unlike in the outflow, \ce{C+} is also involved in the C-chain chemistry in this region, which reduces the \ce{HCO+} abundance.

\paragraph*{Increase in column density relative to the outflow}

The column density of \ce{C2H2} increases by about 3 orders of magnitude due to photochemistry in the outer regions of the disk.
As stated earlier, \ce{C2H2} plays an important role in the chemistry in this region.
Reactions with \ce{SO+}, formed by photoionisation of SO and by \ce{OH + S+}, produce \ce{C2S}, the column density of which increases by two to three orders of magnitude.
Reactions with Si, liberated by photodissociation of the parent SiO, produce \ce{SiC2}, whose column density increases by some 5 orders of magnitude at $10^5$ yr.
SiC is a product of \ce{SiC2 + O}, with the latter produced mostly via the photodissociation of the parents \ce{H2O} and SiO.

These four species (\ce{C2H2}, \ce{C2S}, \ce{SiC2}, and SiC) show large abundances at the inner edge of the disk, within the first few au (marked by the dark orange region in Fig. \ref{fig:coldens-ratio}).
This is caused by the strong temperature dependence of the \ce{H2 + C2H} reaction: its reaction rate depends on temperature as $T^{2.57}$ \citep{Laufer2004}.
This has a significant effect at the large temperatures in this region (between 1100 and 1750 K).
Even though the \ce{Si + C2H2} reaction has a negative temperature dependence ($T^{-0.71}$, \citealt{Canosa2001}), the large increase in \ce{C2H2} abundance leads to a larger \ce{SiC2} and SiC abundance in this region.

The formation of SiN and \ce{SiH4} is linked to Si.
The SiN column density is about an order of magnitude larger than in the outflow at $10^5$ yr. 
Its production is started by \ce{NH3 + Si+}, with the latter produced by the photoionisation of Si or charge exchange with \ce{S+}.
This reaction forms \ce{SiNH2+} which dissociatively recombines with electrons to SiN or recombines to HNSi,  which is then photodissociated into SiN. 
Closer to the midplane, the cation dissociatively recombines with negatively charged dust grains.
The \ce{SiH4} column density is more than 5 orders of magnitude larger at $10^5$ yr.
It is formed by the successive addition of hydrogen to \ce{SiH+} into \ce{SiH5+}, which then reacts with \ce{H2O} or dissociatively recombines with electrons to \ce{SiH4}.
\ce{SiH+} is formed by \ce{Si + H3O+}, with the latter formed by the addition of H to \ce{H2O+}, a product of the photoionisation of \ce{H2O}.

\section{Conclusions}			\label{sect:conclusions}

We presented an updated model of the density and temperature of the disk around \lpup\ by fitting archival ALMA \ce{^{12}CO} and \ce{^{13}CO} $J = 3-2$ lines.
The disk has a surface density profile $\Sigma(r) \sim r^{-1}$ and is warm, with a midplane temperature of 900 K and $T_\mathrm{mid}(r) \sim r^{-0.2}$. 
Unlike \citet{Homan2017}, we do not require a three-staged radial temperature profile or an extremely steep vertical density profile (our $\rho(r) \sim r^{-0.4}$ vs their $\sim r^{-3.1}$.
A more simple parametrisation used commonly for PPDs is able to reproduce the observations.

The chemical model presented in this paper is the first chemical model of a disk around an AGB star, specifically of \lpup's disk.
The physical structure of the disk has a large impact on its chemistry.
All species are constrained to the disk, within $\sim 15$ au radially and $\sim 10$ au vertically. 
This is much smaller than the extent of the parents' envelope in the outflow, which ranges from $30 - 3000$ au radially from the star and is also where the peak abundances of the daughter species are located.

The chemistry within the disk leads to large abundances of C-bearing molecules, a surprising result given its O-rich composition.
Certain COMs are abundantly formed in the midplane.
Unlike in PPDs, these are formed purely via gas-phase chemistry as dust-gas chemistry does not play a role at the high temperatures of \lpup's disk. 
C-chains are readily formed in the outer regions, though caution is needed as the chemical network does not include fully saturated long (more than three C atoms) carbon-chain species.
Two chemical regimes can be distinguished: cosmic-ray induced chemistry in the midplane and photoinduced chemistry in the outer regions.
The abundances of species are significantly increased or decreased with respect to an O-rich outflow chemical model, allowing us to identify potential and unique tracers of the disk such as \ce{H2CO}, \ce{CH3CN}, and \ce{HC3N}.
Moreover, the chemical model can be used to constrain the age of the disk.
The appearance of a gap in abundance in the midplane of SO and \ce{SO2} after $5 \times 10^5$ yr appears to suggest that the disk around \lpup\ is about $10^5$ yr old, as these features were not observed in the data presented by \citet{Kervella2016}.

Further observations are necessary to further constrain the physical structure of the disk. 
In particular, optically thin CO isotopologues are needed to better pin down the disk mass and density structure, while species such as \ce{H2CO}, \ce{HC3N}, \ce{CH3CN}, and \ce{C3H2} can be used as empirical temperature probes.
Targeting the tracers of the midplane and outer region chemistries will provide an essential test of the chemical model predictions.

Our unique chemical model clearly shows the impact that large-scale structures within an AGB outflow can have on its chemistry.
This exploratory work paves the way for a more general study of AGB disks, ranging in density and temperature structures. 
However, note that the 2D model used in this work predicts the chemistry assuming a fixed density structure; coupling chemistry to a hydrodynamical model is necessary to take the effect of dynamical evolution of the chemistry into account.

\section*{Acknowledgements}

We thank Anita Richards for providing us with the reduced ALMA observations, John D. Ilee for help with data analysis, Akke Corporaal for conversations about post-AGB disks, and Wouter Vlemmings for discussions on methanol. 
MVdS thanks Felix Sainsbury-Martinez for advice when scripting.
This article makes use of the following ALMA data: ADS/JAO.ALMA\#2015.1.00141.S. ALMA is a partnership of ESO (representing its member states), NSF (USA) and NINS (Japan), together with NRC (Canada), NSC and ASIAA (Taiwan), and KASI (Republic of Korea), in cooperation with the Republic of Chile. The Joint ALMA Observatory is operated by ESO, AUI/NRAO and NAOJ.
MVdS acknowledges support from the  European Union's Horizon 2020 research and innovation programme under the Marie Skłodowska-Curie grant agreement No 882991 and the Oort Fellowship at Leiden Observatory. 
TD is supported by the Australian Research Council through a Discovery Early Career Researcher Award (DE230100183) and by the Australian Research Council Centre of Excellence for All Sky Astrophysics in 3 Dimensions (ASTRO 3D), through project number CE170100013.
C.W.~acknowledges financial support from the Science and Technology Facilities Council and UK Research and Innovation (grant numbers ST/X001016/1 and MR/T040726/1).
FDC is a Postdoctoral Research Fellow of the Research Foundation - Flanders (FWO), grant number 1253223N.
TC is a PhD fellow of FWO, grant number 1166722N.

\section*{Data Availability}

The data underlying this article will be shared on reasonable request to the corresponding author.




\bibliographystyle{mnras}
\bibliography{chemistry} 




\appendix

\onecolumn
\newpage

\section{ALMA observations} 		\label{app:alma}

Moment 0 maps of the \ce{^{12}CO} and \ce{^{13}CO} $J = 3-2$ lines are presented in \citet{Kervella2016} (their Figs A.1 and B.1). 
Here, we present channel maps of these lines with beam information.

\begin{figure}
\centering
 \includegraphics[width=1\textwidth]{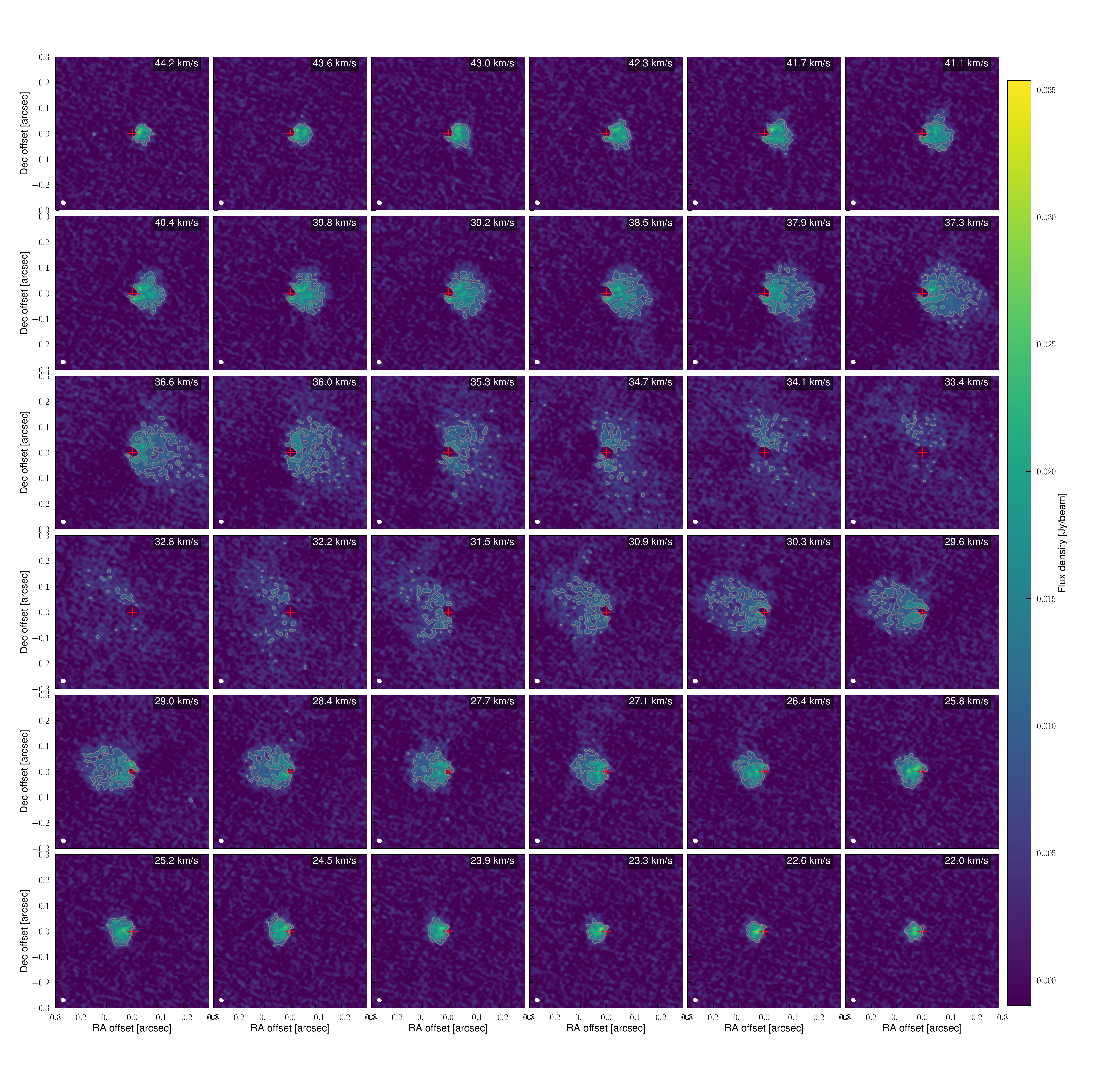}
 \caption{Channel maps of $^{12}$CO $J = 3-2$ emission toward \lpup. 
  The grey contours show flux levels at 3 and 5 times the rms noise. 
  The beam is shown in white in the bottom left hand corners.
 }
 \label{fig:model-mom0}
\end{figure}

\newpage

\begin{figure}
\centering
 \includegraphics[width=1\textwidth]{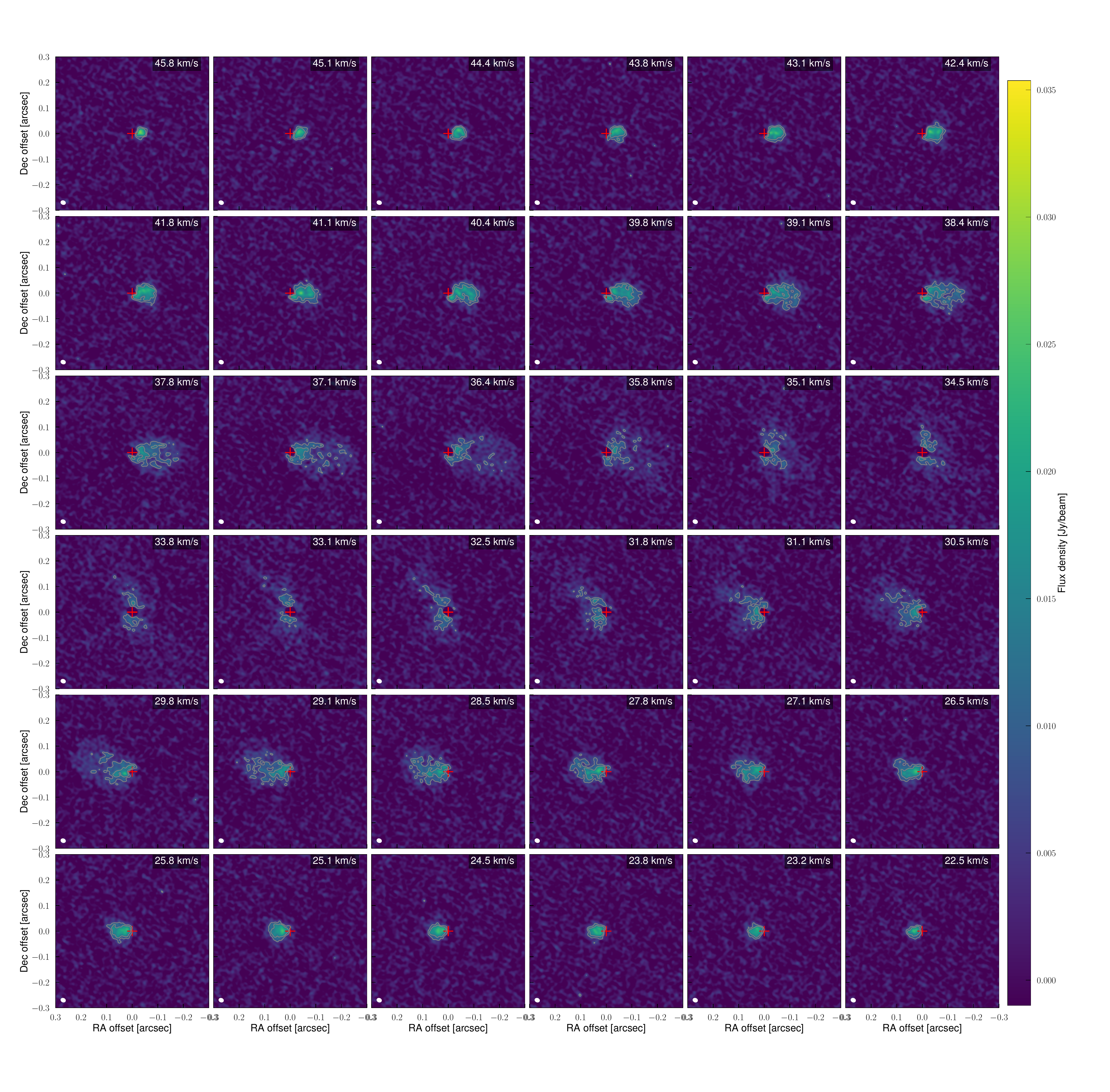}
 \caption{Channel maps of $^{13}$CO $J = 3-2$ emission toward \lpup. 
  The grey contours show flux levels at 3 and 5 times the rms noise. 
  The beam is shown in white in the bottom left hand corners.
 }
 \label{fig:model-mom0}
\end{figure}

\newpage

\section{\textsc{Magritte} modelling results} 		\label{app:magritte}

\begin{figure*}
\centering
 \includegraphics[width=1\textwidth]{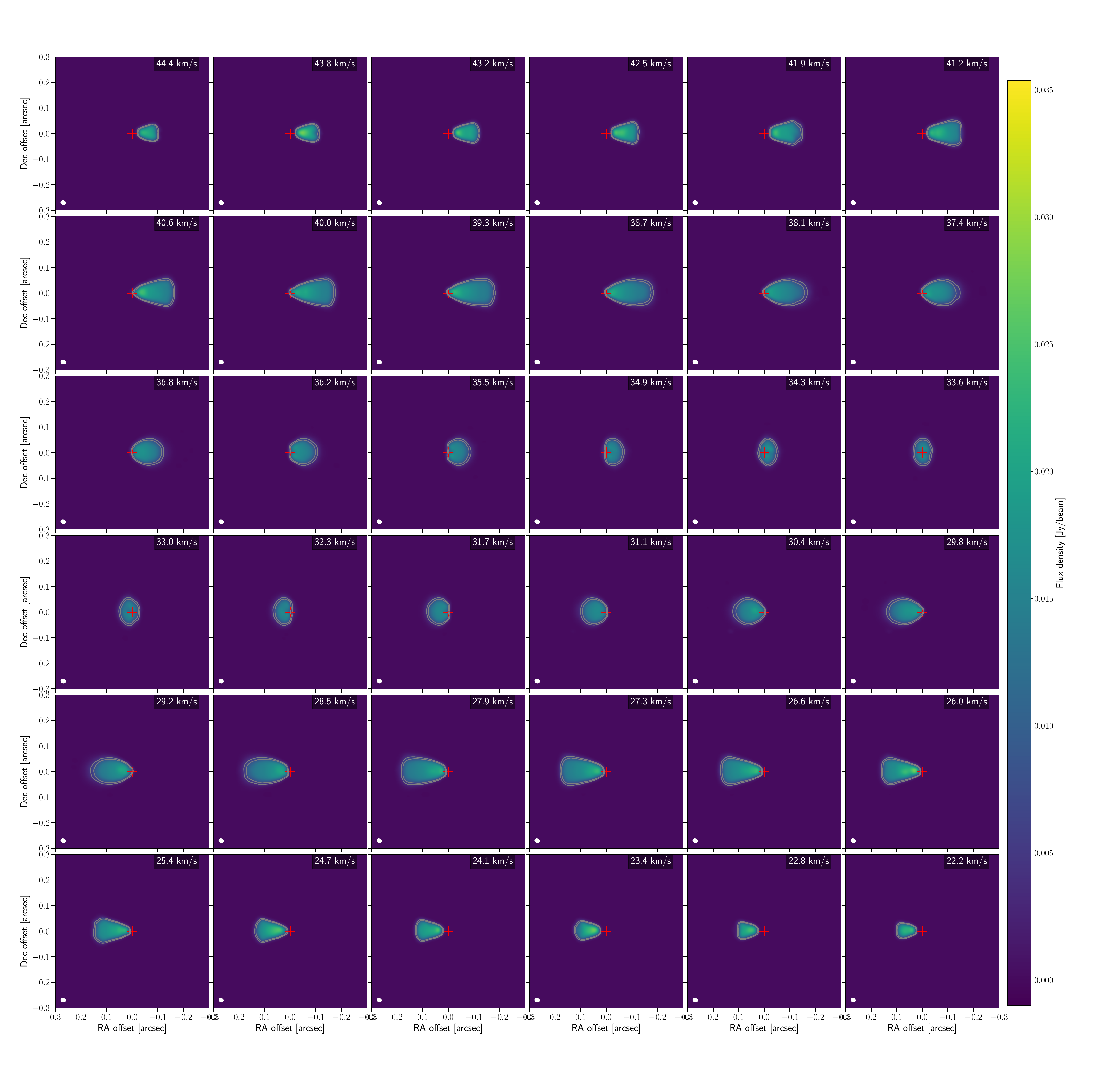}
 \caption{Modelled channel maps of $^{12}$CO $J = 3-2$ emission toward \lpup. 
  The grey contours show flux levels at 3 and 5 times the rms noise. 
  The beam is shown in white in the bottom left hand corners.
  Model parameters are listed in Table \ref{table:modelparams}.
 }
 \label{fig:model-cm-12co}
\end{figure*}

\begin{figure*}
\centering
 \includegraphics[width=1\textwidth]{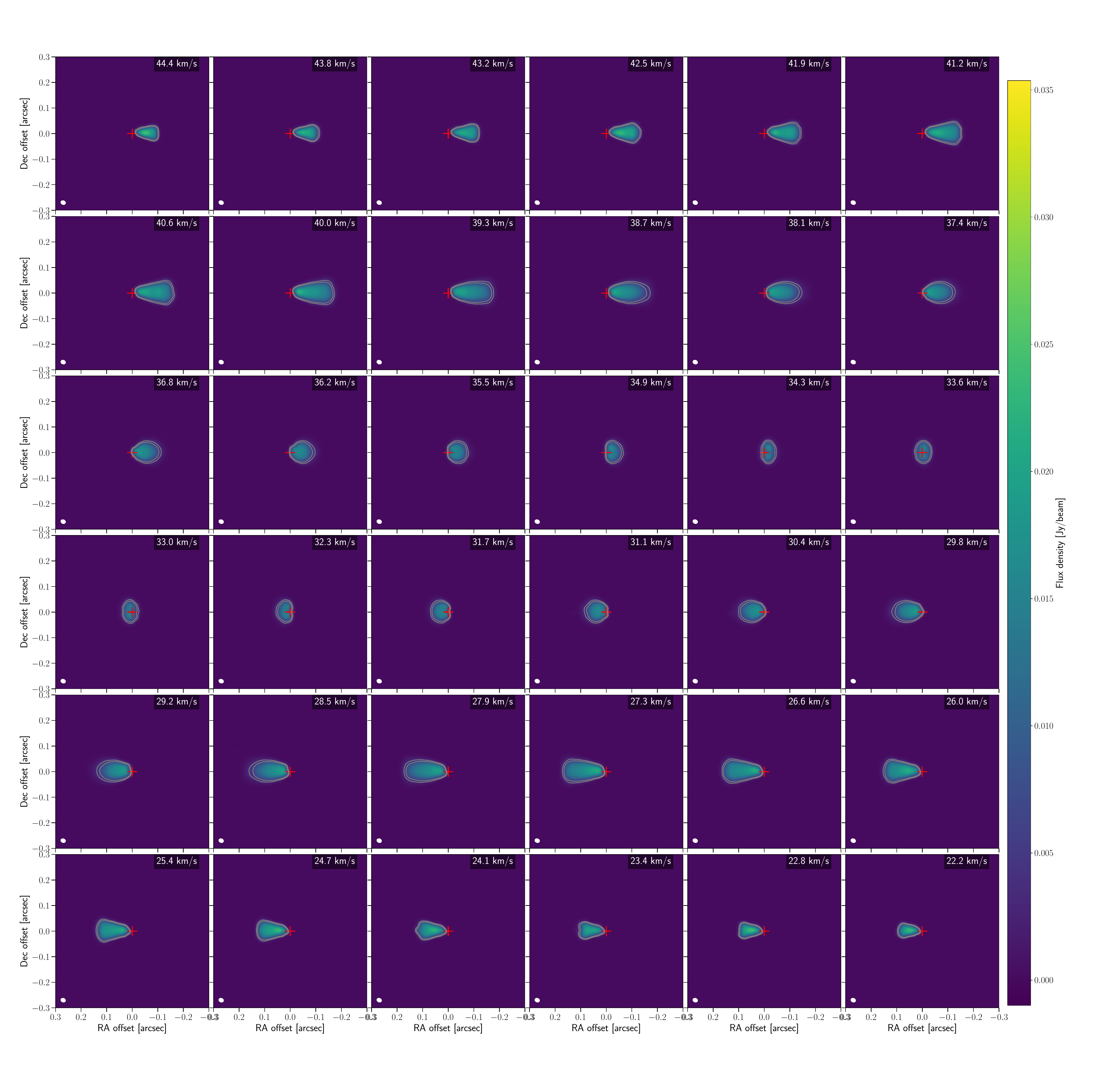}
 \caption{Modelled channel maps of $^{13}$CO $J = 3-2$ emission toward \lpup. 
  The grey contours show flux levels at 3 and 5 times the rms noise. 
  The beam is shown in white in the bottom left hand corners.
  Model parameters are listed in Table \ref{table:modelparams}.
 }
 \label{fig:model-cm-13co}
\end{figure*}

\clearpage

\begin{figure*}
 \includegraphics[width=0.9\textwidth]{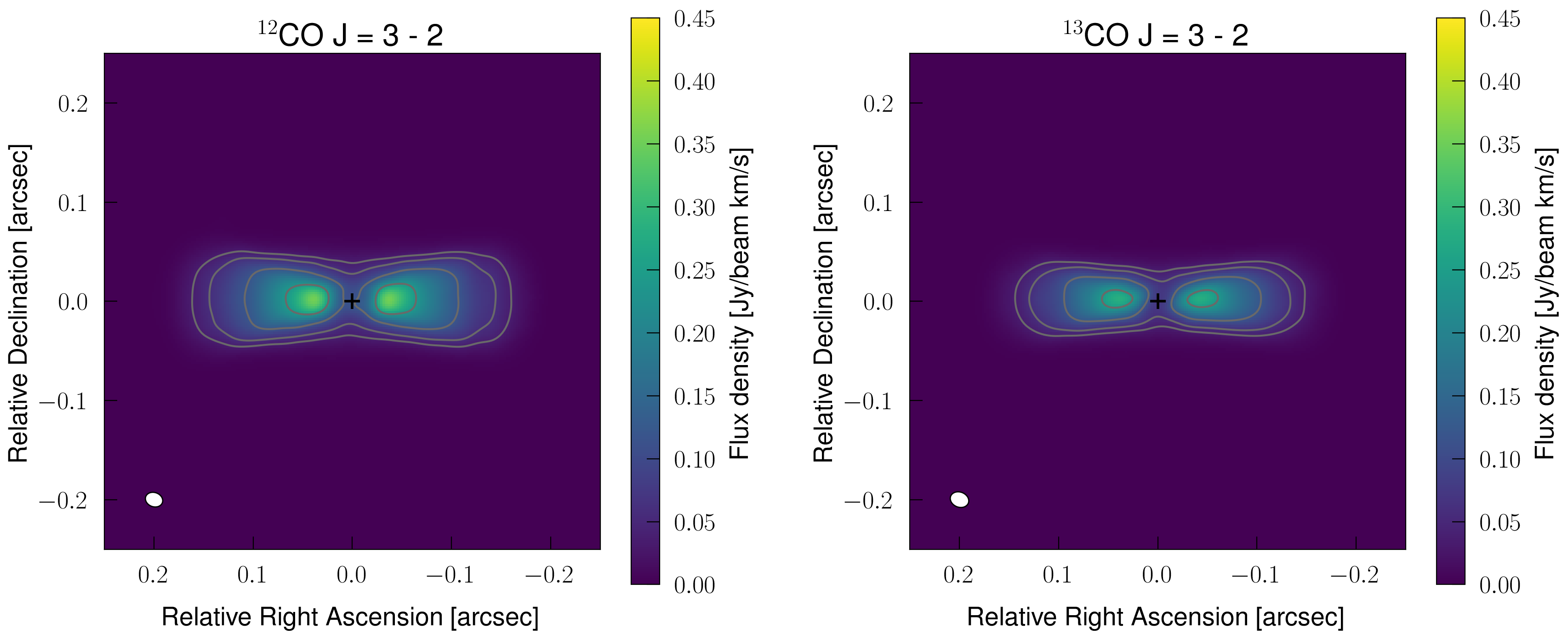}
 \caption{Modelled moment 0 map of $^{12}$CO (left) $^{13}$CO $J = 3-2$ (right). 
  The grey contours show flux levels at 3, 5, 10, 20, and 30 times the rms noise. 
  The beam is shown in white in the bottom left hand corners.
 Model parameters are listed in Table \ref{table:modelparams}.}
 \label{fig:model-mom0}
\end{figure*}

\section{Chemical model results} 		\label{app:chem}

\begin{figure*}
\centering
 \includegraphics[width=0.7\textwidth]{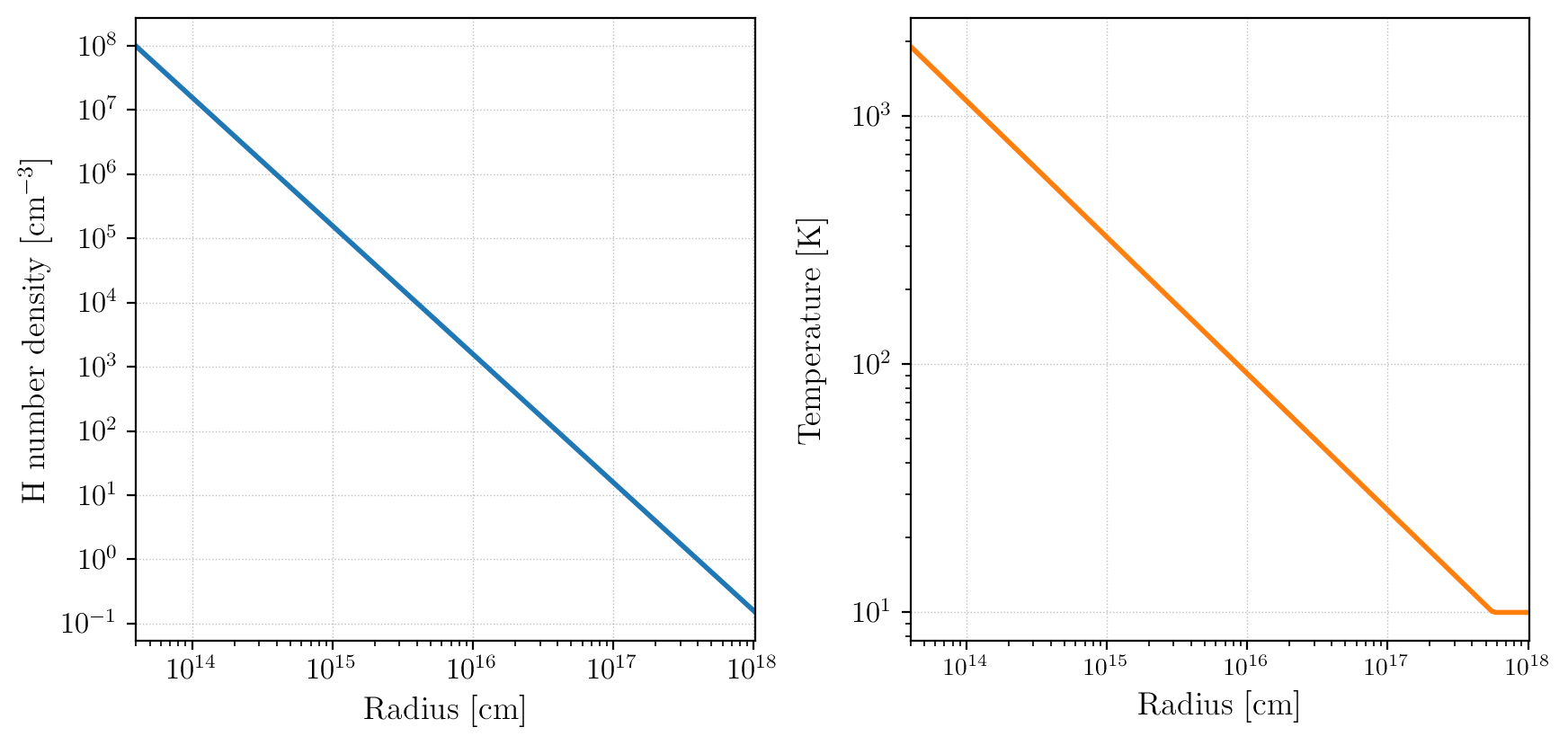}
 \caption{The \ce{H2} number density (left) and temperature (right) assumed throughout the outflow chemical model (Sect. \ref{sect:chem}).
 }
 \label{fig:phys_outflow}
\end{figure*}

\begin{table}
	\caption{Parent species and their initial abundances relative to \ce{H2} for the disk and outflow chemical kinetics models.
	Abundances are derived from observations, as compiled by \citet{Agundez2020}. 
	} 
    \centering
    \begin{tabular}{ l r }
    \hline  
    Species & Initial abundance\\
    \hline  
    \noalign{\smallskip}
     He		& 0.17  \\
     CO		& $3.00 \times 10^{-4}$  \\
     H$_2$O	& $2.15 \times 10^{-4}$  \\
     N$_2$ 	& $4.00 \times 10^{-5}$  \\ 
     SiO 	& $2.71 \times 10^{-5}$  \\ 
     H$_2$S 	& $1.75 \times 10^{-5}$  \\
     SO$_2$ 	& $3.72 \times 10^{-6}$  \\
     SO 		& $3.06 \times 10^{-6}$  \\
     SiS 		& $9.53 \times 10^{-7}$  \\
     NH$_3$ 	& $6.25 \times 10^{-7}$  \\ 
     CO$_2$ 	& $3.00 \times 10^{-7}$  \\   
     HCN 	& $2.59 \times 10^{-7}$  \\  
     PO 		& $7.75 \times 10^{-8}$  \\ 
     CS 		& $5.57 \times 10^{-8}$  \\
     PN 		& $1.50 \times 10^{-8}$  \\
     HCl		& $1.00 \times 10^{-8}$  \\  
     HF	        & $1.00 \times 10^{-8}$  \\    
     Cl	        & $1.00 \times 10^{-8}$  \\    
     Mg	        & $1.00 \times 10^{-8}$  \\    
     Fe	        & $1.00 \times 10^{-8}$  \\    
    \hline 
    \end{tabular}%
    \label{table:model-parents}    
\end{table}

\begin{figure*}
\centering
 \includegraphics[width=1\textwidth]{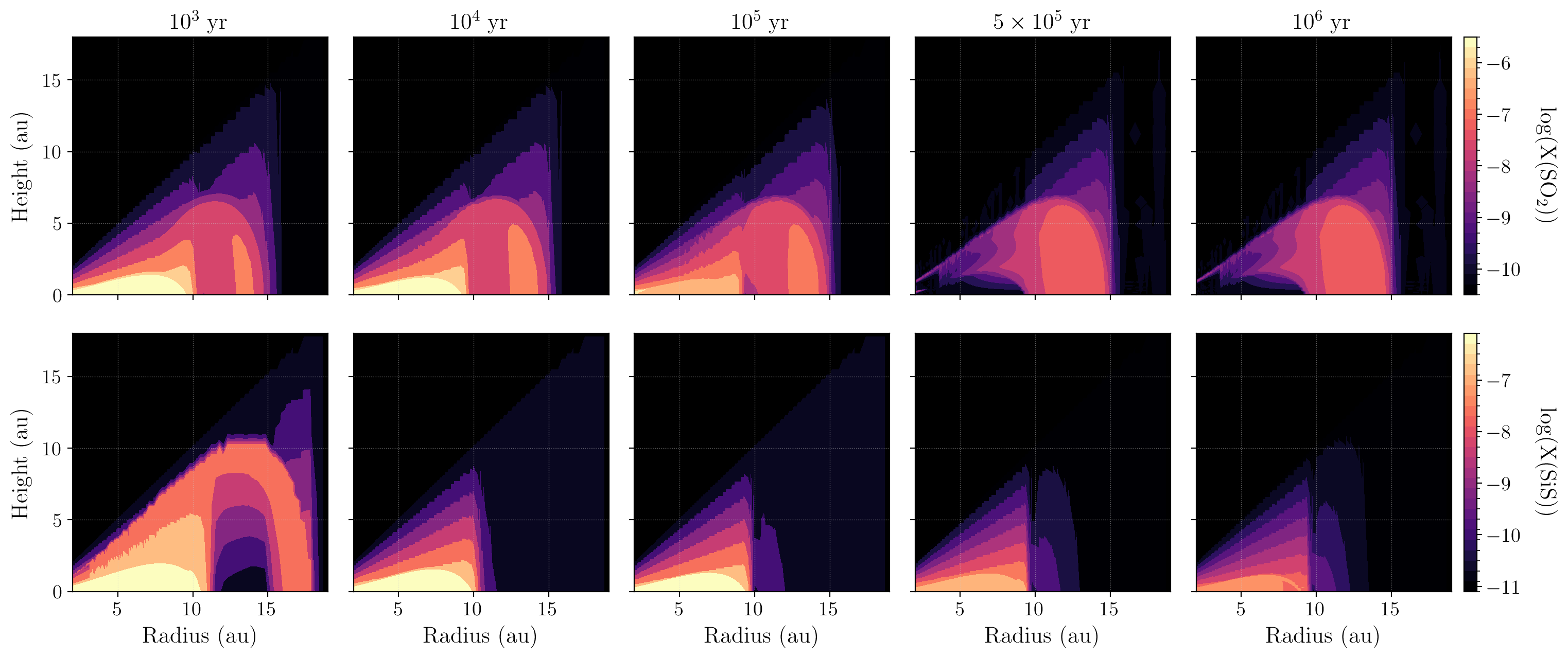}
 \caption{Fractional abundance of \ce{SO2} (top row) and SiS (bottom row) throughout the disk as predicted by the chemical model.
 The colour map shows the logarithm of the abundance w.r.t. \ce{H2}.
 The different columns show the results after different times, increasing from left to right.
 }
 \label{fig:SO2_SiS}
\end{figure*}

\begin{figure*}
\centering
 \includegraphics[width=1\textwidth]{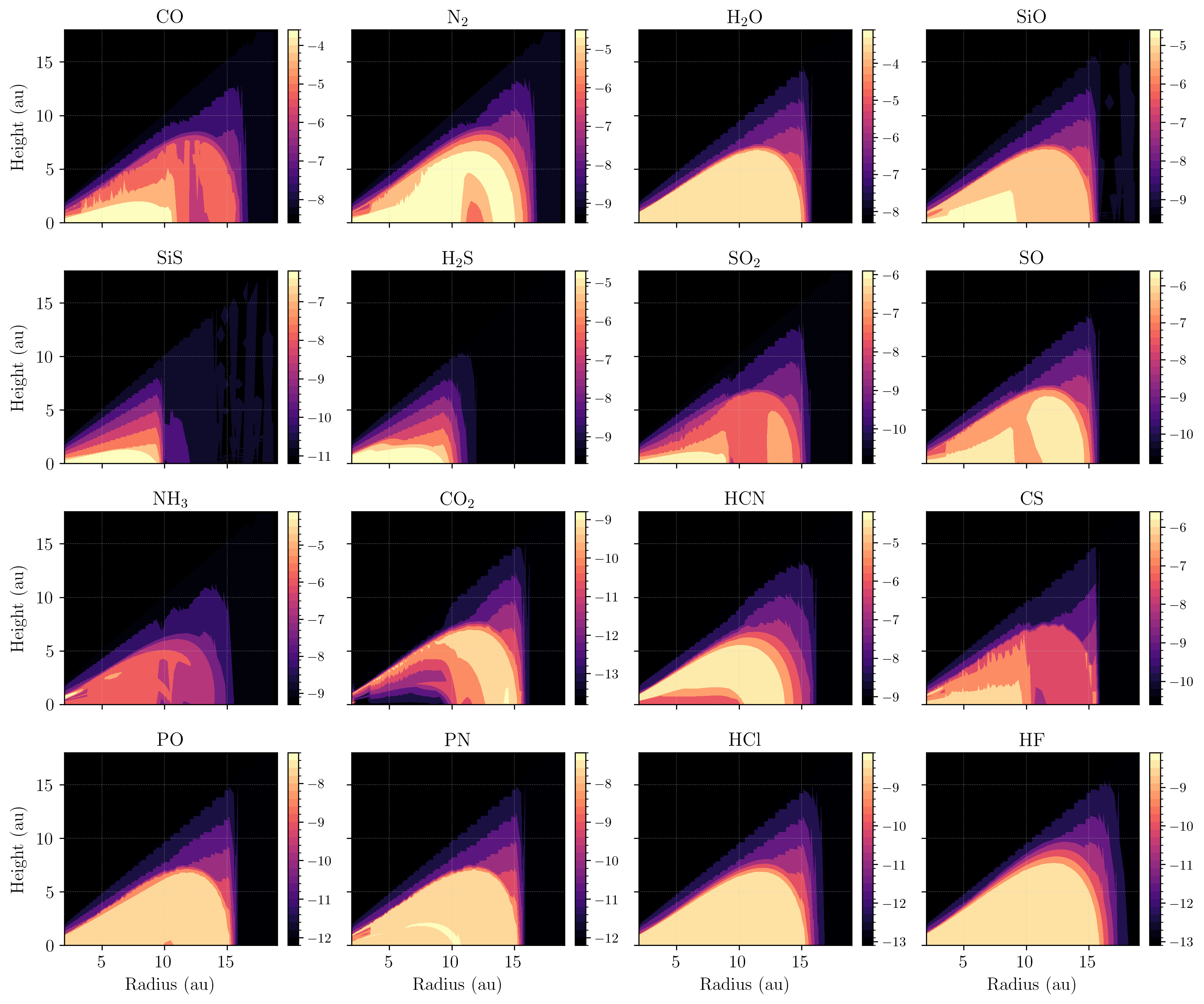}
 \caption{Fractional abundances of the parent species throughout the disk as predicted by the chemical model of the parent species after $10^5$ yr.
 The colour map shows the logarithm of the abundance w.r.t. \ce{H2}.
 Note that each colour map has a different dynamic range.
 }
 \label{fig:allparents}
\end{figure*}

\begin{figure*}
\centering
 \includegraphics[width=0.8\textwidth]{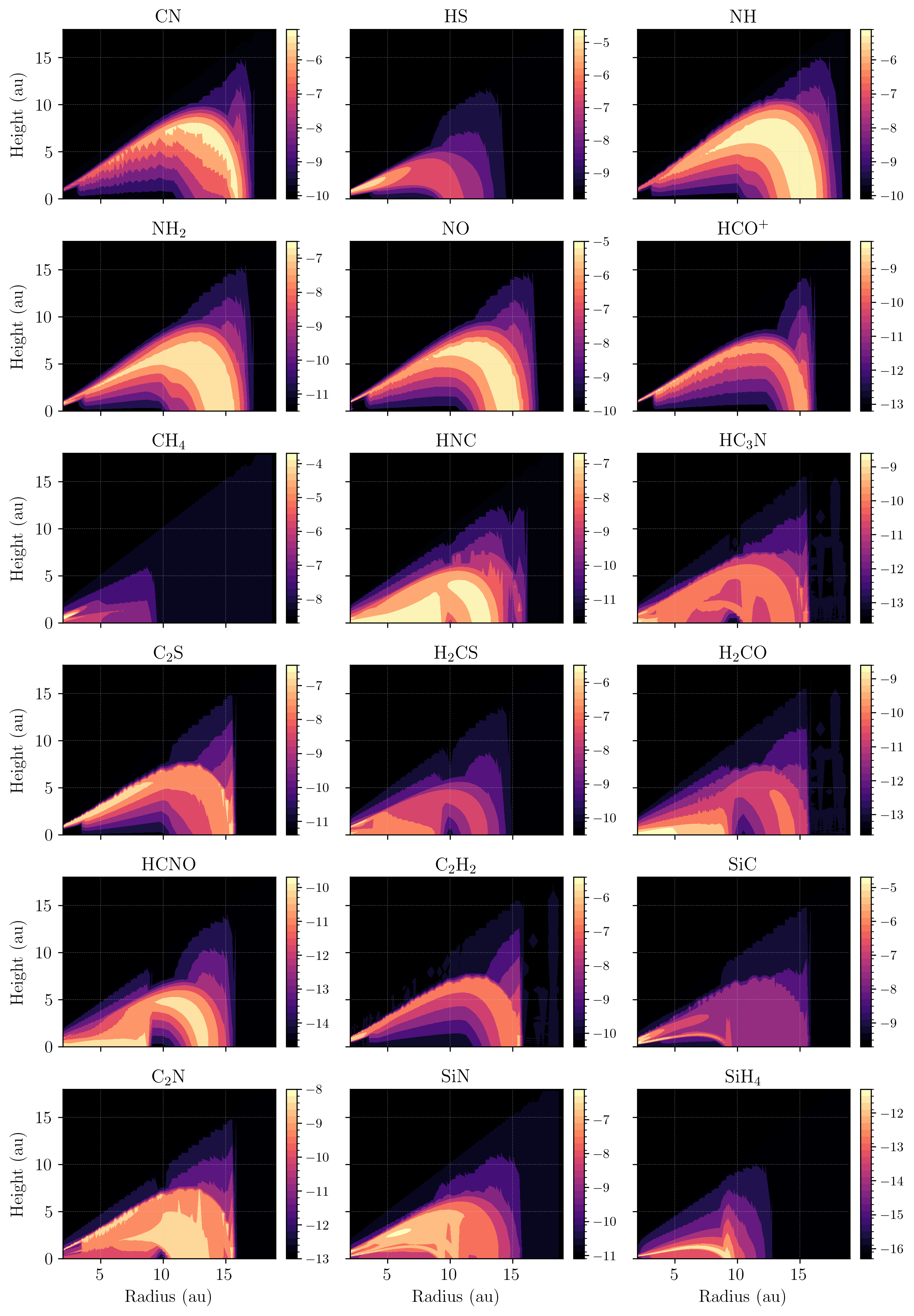}
 \caption{Fractional abundances of some daughter species shown in Fig. \ref{fig:coldens-ratio} throughout the disk as predicted by the chemical model of the parent species after $10^5$ yr.
 The colour map shows the logarithm of the abundance w.r.t. \ce{H2}.
 Note that each colour map has a different dynamic range.
 }
 \label{fig:daughters_ratio}
\end{figure*}

\begin{figure*}
\centering
 \includegraphics[width=1\textwidth]{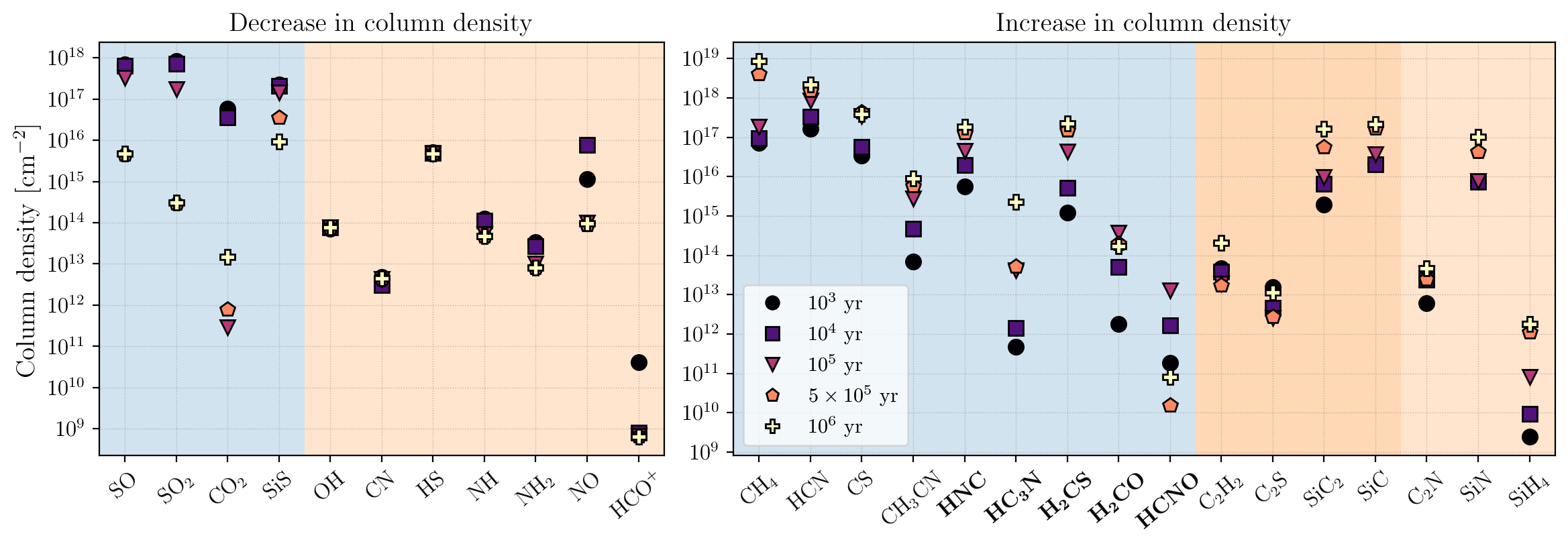}
 \caption{Disk column density for species that show a decrease (left panel) and increase in column density relative to the outflow (right panel) of more than an order of magnitude at any time in the disk chemical model.
 Different markers and colours show the ratio at different times for the disk model.
 The shaded regions group species originating from different types of chemistry.
 Blue: midplane (cosmic ray) chemistry, orange: outer edge (photo) chemistry, dark orange: hot inner edge.
 Boldfaced labels denote species that are formed via cosmic ray and photochemistry.
  }
 \label{fig:coldens-value}
\end{figure*}


\bsp	
\label{lastpage}
\end{document}